\documentclass[11pt,a4paper]{article}
\usepackage{german}
\usepackage{times}
\usepackage{amsmath}
\usepackage{rotating}
\usepackage{epsfig}
\usepackage{euler}
\usepackage{a4}
\newcommand{\E}{\bar{E}}
\usepackage[hang]{footmisc}
\begin{document}

\title{\textbf{,,\emph{Es lebe die Unverfrorenheit\,!}``} \\
Albert Einstein und die Begr"undung der Quantentheorie\footnote{%
Erschienen in: Herbert Hunziker (Hrsg.) \emph{Der jugendliche 
Einstein und Aarau} (Birkh"auser Verlag, Basel, 2005)}}

\author{Domenico Giulini            \\
        Universit"at Freiburg       \\
        Physikalisches Institut     \\
        Hermann-Herder-Stra"se 3    \\
        79104 Freiburg}
\date{}

\maketitle

\begin{abstract}
\noindent
Am 14.\,Dezember des Jahres 1900 berichtete Max Planck der Deutschen 
Physikalischen Gesellschaft "uber seine physikalische Interpretation 
einer harmlos aussehenden, von ihn selbst zuvor aufgestellten Formel, 
die das spektrale Verhalten der sogenannten W"armestrahlung beschreibt. 
Ma"sgeblich durch das Eingreifen Albert Einsteins entwickelte sich 
daraus im folgenden Vierteljahrhundert eine fundamentale Krise der 
Physik, die dann in einer wissenschaftlichen Revolution gr"o"sten 
Ausma"ses m"undete: der Quantentheorie. 

Die Quantentheorie entwickelte sich von Anfang an diametral gegen die 
Intentionen ihrer Sch"opfer. F"ur Planck bedeutete sie -- trotz gr"o"ster 
"au"serer Anerkennungen -- das vollst"andige Scheitern eines langj"ahrigen 
Forschungsprogramms, f"ur Einstein letztlich eine Absage an seine 
wissenschaftlichen Grund"uberzeugungen. Wir schildern die Hintergr"unde 
dieser seltsamen Entwicklung und beleuchten damit die begriffliche 
Seite physikalischer (und allgemein naturwissenschaftlicher) Forschung, 
die gemeinhin stark untersch"atzt wird.           
\end{abstract}

\noindent
Um die Rolle zu verstehen, die Albert Einstein bei der Entwicklung 
der Quantentheorie gespielt hat, m"ussen wir uns zun"achst die 
vorangegangenen Leistungen Plancks vergegenw"artigen, die ihn zur 
Aufstellung seiner ber"uhmten Strahlungsformel gef"uhrt haben. 
Mit dieser gelang ihm die vollst"andige \emph{quantitative} 
Aufkl"arung des Ph"anomens der \emph{W"armestrahlung}, die ihm 
den Nobelpreis des Jahres 1918 einbrachte: "`als Anerkennung des 
Verdienstes, das er sich durch seine Quantentheorie um die Entwicklung 
der Physik erworben hat"'. Zu diesem Zeitpunkt lag die eigentliche 
Tat schon mehr als 17 Jahre zur"uck.  Genauer ist sie auf den 
14.\,Dezember des Jahres 1900 zu datieren. Davon wird weiter unten 
die Rede sein. 

Etwas weniger bekannt ist die Tatsache, da"s diese wissenschaftliche 
Gro"stat Plancks gleichzeitig auch die restlose Zerschlagung seines 
langj"ahrigen, akribisch vorbereiteten und meisterhaft durchgef"uhrten 
Forschungsprogramms bedeutete, das in einer tief anti-atomistischen, 
an absoluten Gesetzm"a"sigkeiten sich orientierenden Naturauffassung 
wurzelt. In der Verfolgung dieser Ideale legt Planck den Grundstein 
zur Quantentheorie, die dem konsequenten Atomismus zum endg"ultigen 
Durchbruch verhilft und dem Element des Zufalls eine fundamentale 
Bedeutung innerhalb des Gef"uges physikalischer Gesetzm"a"sigkeiten 
zuweist. Hauptmotor dieser Entwicklung, die den Planckschen Vorstellungen 
diametral entgegenlief, war Albert Einstein. Hartn"ackig und mitunter  
unverfroren\footnote{Das Zitat des Titels entstammt einem Brief
(\cite{Einstein-CW}, Band\,1, Dokument\,127) Einsteins vom 12.\,Dezember
1901 an seine damalige Freundin und sp"atere Frau Mileva Maric, 
in dem er seine Courage in einer privaten Angelegenheit zu einer Art 
Lebensmotto erhob und kommentierte: "`Es lebe die Unverfrorenheit! 
Sie ist mein Schutzengel in dieser Welt."'} bestand er auf der 
restlosen Kl"arung 
der begrifflichen Grundlagen und Konsequenzen der Planckschen Theorie.
Mit seiner Lichtquantenhypothese erkl"arte er nicht nur den 
photoelektrischen Effekt, sondern legte den eigentlich 
revolution"aren Kern dieser Theorie frei und provozierte so 
ma"sgeblich eine tiefe Krise, die 20 Jahre sp"ater in der Formulierung 
der Quantenmechanik m"undete. Etwas "ubertreibend, aber im Kern doch 
zutreffend, kann man sagen, da"s Einstein der einzige war, der 
die Plancksche Theorie wirklich ernst nahm -- so ernst, da"s die 
Konsequenzen sich schlie"slich auch gegen seine Grund"uberzeugungen 
richteten.

\section{Plancks Programm}
Planck hatte sich schon in jungen Jahren ein ehrgeiziges 
Forschungsprogramm zurechtgelegt. Er wollte den sogenannten 
2.\,Hauptsatz\footnote{Der 1.\,Hauptsatz ist der Satz "uber die 
Erhaltung der Energie. Der 2.\, Hauptsatz betrifft nicht die 
Energie, sondern eine andere Zustandsgr"o"se, genannt \emph{Entropie}.
Er besagt in der "alteren, Planck n"aherliegenden Formulierung, da"s 
die Entropie zeitlich nicht abnimmt. In der modernen, von Planck 
zun"achst bek"ampften statistischen Interpretation der Entropie, 
ist diese ein Ma"s f"ur die "`Unordnung"'. Genauer gesagt ist die 
Entropie ein (logarithmisches) Ma"s f"ur die Anzahl der Mikrozust"ande, 
die einen makroskopisch definierten Zustand realisieren (siehe dazu 
Anhang\,\ref{sec:AnhangPlanck}). 
Der 2.\,Hauptsatz besagt in dieser Interpretation, da"s die Entropie 
\emph{im Zeitmittel} nicht abnimmt (statistische Schwankungen, in denen 
die Entropie vor"ubergehend kurz abnimmt, sind also erlaubt). 
Der 2.\,Hauptsatz regelt die Irreversibilit"at gewisser Prozesse. 
Das sind dann solche, bei denen die Entropie zunimmt.} der 
Thermodynamik mit Hilfe der Theorie elektromagnetischer Vorg"ange 
streng begr"unden. Dies geschah aus einer Opposition zu den Vertretern 
des Atomismus, die in den Gesetzen der Thermodynamik lediglich 
statistische Gesetzm"a"sigkeiten einer sonst regellosen Bewegung sehr 
vieler Molek"ule sehen wollten, w"ahrend Planck fest an eine strenge 
Gesetzlichkeit ohne statistische Ausnahmen glaubte.  
In einer Jugendarbeit aus dem Jahre 1884 schreibt 
der 24-j"ahrige selbstbewu"st (\cite{Planck-GW}, Band\,I, 
Dokument Nr.\,4, pp.\,162-163):
\begin{quote}
"`Der zweite Hauptsatz der mechanischen W"armetheorie
consequent durchgef"uhrt, ist unvertr"aglich mit der
Annahme endlicher Atome. Es ist daher vorauszusehen,
da"s es im Laufe der weiteren Entwicklung der Theorie
zu einem Kampfe zwischen diesen beiden Theorien
kommen wird, der einer von ihnen das Leben kostet."'
\end{quote}
Zwei Zeilen weiter l"a"st er wenig Zweifel dar"uber, 
welche der Theorien seiner Meinung und Hoffnung nach das 
Leben wird lassen m"ussen:
\begin{quote}
"`... indessen scheinen mir augenblicklich verschiedenartige
Anzeichen darauf hinzudeuten, da"s man trotz der
bisherigen Erfolge der atomistischen Theorie sich
schlie"slich doch einmal zu einer Aufgabe derselben
und zur Annahme einer continuierlichen Materie wird
entschlie"sen m"ussen."'
\end{quote}
Zu dieser Zeit war der junge Planck ein erkl"arter Anti-Atomist. 
Sein Plan war, zu versuchen, die thermodynamischen Gesetze nicht 
"uber eine Mechanik elementarer Konstituenten (Atome, Molek"ule) 
zu begr"unden, sondern mit Hilfe der Gesetze der Elektrodynamik, 
die mit rein kontinuierlichen, im Raum verteilten Gr"o"sen operiert. 
In seiner Antrittsrede anl"a"slich seiner Aufnahme in die Preu"sische 
Akademie der Wissenschaften im Jahre 1894 erkl"arte er
(\cite{Planck-GW}, Band \,III, Dokument Nr.\,122, p.\,3):
\begin{quote}
"`Es hat sich neuerdings in der physikalischen Forschung auch 
das Bestreben Bahn gebrochen, den Zusammenhang der 
Erscheinungen "uberhaupt gar nicht in der Mechanik zu suchen [..].
Ebenso steht zu hoffen, da"s wir auch "uber diejenigen 
elektrodynamischen Prozesse, welche direkt durch die Temperatur
bedingt sind, wie sie sich namentlich in der W"armestrahlung 
"au"sern, n"ahere Aufkl"arung erfahren k"onnen, ohne erst den 
m"uhsamen Umweg durch die mechanische Deutung der Elektrizit"at 
nehmen zu m"ussen."'
\end{quote}
Planck glaubte also an die M"oglichkeit, die Gesetze der 
Thermodynamik, namentlich den 2.\,Hauptsatz, als strenge 
Folge bekannter elektromagnetischer Gesetze zu 
verstehen.\footnote{Aus heutiger Sicht ist diese Hoffnung 
schwer verst"andlich, da die Gesetze der Elektrodynamik genauso wie 
die Gesetze der Mechanik \emph{invariant unter Bewegungsumkehr} sind. 
Das bedeutet, da"s mit jeder den Gesetzen gen"ugenden Bewegung die 
entsprechend zeitlich r"uckl"aufige Bewegung wieder eine m"ogliche 
Bewegung im Sinne der Gesetze ist. Aus dieser mathematischen 
Tatsache folgt zwingend die Unm"oglichkeit eines Beweises "uber 
die ausnahmslose zeitliche Zunahme einer Zustandsgr"o"se, wie etwa 
der Entropie. Nur unter \emph{zus"atzlichen} Annahmen, die immer 
Einschr"ankungen an die Anfangsbedingungen beinhalten, k"onnen 
solche Beweise funktionieren. Auch Planck wird sp"ater bei seiner 
`Ableitung' der Wienschen Strahlungsformel eine solche Annahme in 
etwas versteckter Form machen (durch seine "`Hypothese der 
nat"urlichen Strahlung"'), was f"ur die hier zu besprechenden 
Entwicklungen aber nicht weiter relevant ist. Noch schwerer 
verst"andlich wird das Festhalten Plancks an dieser Hoffnung durch 
den Hinweis, da"s Planck das eben skizzierte Argument sicherlich 
kannte, n"amlich durch den Mathematiker Ernst Zermelo, der in den 
Jahren 1894-1897 sein Assistent war und dar"uber einiges publiziert 
hat.} Dieser sollte aus allgemeinsten Prinzipien ableitbar sein, 
entsprechend seiner wissenschaftlichen Disposition, die er in seinem 
sp"aten, pers"onlich gehaltenen Artikel "`Zur Geschichte der 
Auffindung des physikalischen Wirkungsquantums"' aus dem Jahre 1943 
so charakterisierte
(\cite{Planck-GW}, Band\,III, Dokument\,141, p.\,255):
\begin{quote}
 "`Was mich in der Physik von jeher vor allem interessierte,
waren die gro"sen allgemeinen Gesetze, die f"ur s"amtliche
Naturvorg"ange Bedeutung besitzen, unabh"angig von den
Eigenschaften der an den Vorg"angen beteiligten K"orper."'
\end{quote} 

\section{Fr"uhe Strahlungstheorie}
Man denke sich einen Hohlraum, der vollst"andig durch W"ande 
umschlossen ist, etwa das Innere eines Ofens. Bringt man die 
W"ande auf eine konstante Temperatur\footnote{Aus bestimmten 
Gr"unden benutzen Physiker lieber die sogenannte absolute 
Temperaturskala, auf der die Temperatur nicht in Grad Celsius, 
sondern in Grad Kelvin angegeben wird. Beide Skalen unterscheiden 
sich um den konstanten Betrag von $273{,}15$, d.h. $X$ Grad Celsius 
entsprechen $X+273{,}15$ Grad Kelvin. Null Grad Kelvin, also $-273{,}15$ 
Grad Celsius, bildet eine absolute untere Grenze f"ur alle 
erreichbaren Temperaturen, die unter keinen Umst"anden unterschritten 
werden kann.} $T$, so wird sich nach einiger Zeit im Hohlraum eine 
bestimmte Konfiguration elektromagnetischer Strahlung einstellen,
die sogenannte W"armestrahlung. Diese wird aus elektromagnetischen 
Wellen aller Frequenzen mit unterschiedlichen Intensit"aten bestehen.
Zwischen Strahlung und der die W"ande bildenden Materie wird nach 
einiger Zeit ein thermodynamisches Gleichgewicht bestehen. 
Einzig wesentliche Voraussetzung f"ur die Existenz eines stabilen 
Gleichgewichtszustandes ist die Annahme, da"s die Materie (oder 
zumindest Anteile davon) in \emph{allen} Frequenzbereichen mit der 
Strahlung wechselwirkt, also Strahlung aller Frequenzen emittieren 
und absorbieren kann. Mit Hilfe dieser Annahme folgerte Gustav 
Kirchhoff bereits 1859 die Existenz einer \emph{universellen} Funktion 
$\rho(\nu,T)$ f"ur die spektrale Energieverteilung der Strahlung. 
Diese gibt an, wieviel Energie in Form von elektromagnetischen 
Wellen der Frequenz $\nu$ (genauer: in einem kleinen Frequenzintervall 
$[\nu,\nu+d\nu]$ um den Wert $\nu$) in einem Einheitsvolumen 
(z.B. Kubikzentimeter) des Hohlraumes enthalten ist, wenn die W"ande 
auf die Temperatur $T$ aufgeheizt wurden. Da"s diese Funktion 
"`universell"' ist, bedeutet, da"s sie \emph{nicht} von der genaueren 
Beschaffenheit der W"ande abh"angt, also nicht von ihrer Form oder 
ihrem Material. Egal, ob die W"ande aus Kupfer, Uran, Keramik oder 
sonstwas bestehen, immer wird sich bei vorgegebener Temperatur ein 
und dieselbe spektrale Energieverteilung von W"armestrahlung 
einstellen. Darin liegt die nichttriviale Einsicht Kirchhoffs. 
Daraus entsteht nun die  theoretische Aufgabe, diese universelle 
Funktion aus den bekannten Gesetzen der Thermodynamik und 
Elektrodynamik zu bestimmen. Man beachte, da"s diese Aufgabe nur 
wegen der Universalit"at l"osbar erscheint, da dadurch die Kenntnis 
komplizierter Materialeigenschaften sowie deren (zum damaligen 
Zeitpunkt gr"o"stenteils unbekannter) Einfl"usse auf die 
Wechselwirkung zwischen Material und Strahlung nicht vorausgesetzt 
werden m"ussen.

Durch weitere, raffiniertere thermodynamische "Uberlegungen 
konnte Wilhelm Wien 1893 zeigen, da"s die Funktion $\rho(\nu,T)$
aus dem Produkt der dritten Potenz der Frequenz $\nu$ und einer 
Funktion $f$ bestehen mu"s, die jetzt nur noch von \emph{einer}
Variablen abh"angt, n"amlich dem Quotienten der Frequenz und der 
Temperatur\footnote{Zwar treten im Argument der Funktion sowohl 
die Frequenz als auch die Temperatur $T$ auf, aber nur als 
Quotient $\nu/T$. Dieser Quotient ist die \emph{eine} Variable, 
von der $f$ alleine abh"angt.}. Es mu"s also gelten:
\begin{equation}
\label{eq:Wien1}
\rho(\nu,T)=\nu^3f(\nu/T)\,.
\end{equation}
Der Fortschritt dieser Einsicht Wiens besteht also in der Reduktion 
des Problems auf die Bestimmung einer Funktion mit nur \emph{einer} 
anstatt zwei unabh"angigen Variablen. Bestimmt man $f$, so ist damit 
nach (\ref{eq:Wien1}) auch $\rho(\nu,T)$ bekannt. Au"serdem folgen aus 
(\ref{eq:Wien1}) auch ohne Kenntnis der Funktion $f$ bereits erste, 
experimentell pr"ufbare Konsequenzen, die gl"anzend best"atigt wurden. 
So ergibt sich einerseits das sogenannte Wiensche Verschiebungsgesetz, 
welches besagt, da"s die Frequenz, bei der die spektrale 
Energieverteilung ihr Maximum hat, proportional mit der Temperatur 
w"achst. Ebenso ergibt sich, da"s die gesamte, "uber alle Frequenzen 
summierte Energieabstrahlung mit der vierten Potenz der Temperatur 
anw"achst. Dies bezeichnet man als das Stefan-Boltzmannsche Gesetz. 

Wie gesagt, bestand die eigentliche Aufgabe nun in der Bestimmung 
der einen Funktion $f$. Durch weitere Anwendung fundamentaler 
Prinzipien sollte dies schlie"slich ohne allzu gro"sen Aufwand gelingen --
so dachten die Physiker zwischen 1893 und 1900. Doch erwies sich 
diese Aufgabe "uberraschenderweise als fast unl"osbar. 
R"uckschauend aus dem Jahre 1913 charakterisierte Einstein die 
Situation so (\cite{Einstein-CW}, Band\,4, Dokument Nr.\,23, p.\,562): 
\begin{quote}
"`Es w"are erhebend, wenn wir die Gehirnsubstanz auf eine
Waage legen k"onnten, die von den theoretischen Physikern 
auf dem Altar dieser universellen Funktion $f$ hingeopfert
wurde; und es ist diesen grausamen Opfers kein Ende abzusehen!
Noch mehr: auch die klassische Mechanik fiel ihr zum Opfer,
und es ist nicht abzusehen, ob Maxwells Gleichungen der 
Elektrodynamik die Krisis "uberdauern werden, welche diese 
Funktion $f$ mit sich gebracht hat."'
\end{quote}

Doch zur"uck zum Geschehen. Aus "Uberlegungen, die man eher als 
"`educated guessing"' bezeichnen kann, schl"agt Wien eine 
einfache Exponentialfunktion f"ur $f$ vor, die dann im Verbund 
mit (\ref{eq:Wien1}) zum sogenannten \emph{Wienschen Strahlungsgesetz} 
f"uhrt ($\exp$ bezeichnet im folgenden die Exponentialfunktion): 
\begin{equation}
\label{eq:Wien2}
\rho(\nu,T)=a\nu^3\exp\left[-\frac{b\nu}{T}\right]\,,
\end{equation}
wobei $a$ und $b$ Konstanten sind, die es noch zu bestimmen
gilt. 

Zahlreiche Experimente schienen ausnahmslos diese Form der 
spektralen Energieverteilung zu best"atigen (dies blieb der Fall bis 
etwa Mitte 1900). "Uberzeugt von seiner Richtigkeit stellt sich daher 
Planck die Aufgabe, das Wiensche Strahlungsgesetz aus ersten 
Prinzipien theoretisch abzuleiten. Als "`Prinzipienlieferant"' akzeptiert 
er vornehmlich die Elektrodynamik und die Thermodynamik und hier an 
erster Stelle den 2.\,Hauptsatz "uber die Zunahme der Entropie. 
Nach langen M"uhen gelingt ihm schlie"slich im Jahre 1899 eine 
Ableitung von (\ref{eq:Wien2}). Er res"umiert stolz
(\cite{Planck-GW}, Band\,I, Dokument Nr.\,34, p.\,597):
\begin{quote}
"`Ich glaube hieraus schlie"sen zu m"ussen, da"s die gegebene
Definition der Strahlungsentropie und damit auch das Wiensche
Energieverteilungsgesetz eine notwendige Folge der Anwendung des
Principes der Vermehrung der Entropie auf die elektromagnetische
Strahlungstheorie ist und da"s daher die Grenzen der G"ultigkeit
dieses Gesetzes, falls solche "uberhaupt existieren, mit denen des
zweiten Hauptsatzes der W"armetheorie zusammenfallen."'
\end{quote}
Ironischerweise sind es Experimentalphysiker (Lummer und Pringsheim), 
die den Theoretiker Planck in einer Ver"offentlichung des gleichen Jahres,
die der experimentellen "Uberpr"ufung des Wienschen Strahlungsgesetzes 
gewidmet ist, h"oflich darauf hinweisen, da"s hier ein logisch unzul"assiger 
Umkehrschlu"s vorliegt (\cite{Lummer-Pringsheim}, p.\,225): 
\begin{quote}
"`Herr Planck spricht es aus, da"s dieses [d.h. (\ref{eq:Wien2})] 
Gesetz eine nothwendige Folge der Anwendung des Principes der Vermehrung 
der Entropie auf die elektromagnetische Strahlung ist, und da"s daher 
die Grenzen seiner G"ultigkeit, falls solche "uberhaupt existieren, 
mit denen des zweiten Hauptsatzes der W"armetheorie zusammenfallen. 
Soviel uns scheint, w"are die Planck'sche Theorie erst zwingend, wenn 
wirklich nachgewiesen werden kann, da"s \emph{jede} von obiger Gleichung 
abweichende Form zu einem Ausdruck der Entropie f"uhrt, der dem 
Entropiegesetz widerspricht."'
\end{quote} 
Planck hatte n"amlich keineswegs gezeigt, da"s das Wiensche Gesetz eine 
logische Folge des 2.\,Hauptsatzes der Thermodynamik ist, sondern nur,
da"s es dem 2.\,Hauptsatz nicht widerspricht. Trotz dieses logischen 
Lapsus' ist die von Planck verwendete Methode bemerkenswert. Da sie 
charakteristisch f"ur das Vorgehen eines theoretischen Physikers 
ist, soll sie hier etwas ausf"uhrlicher beschrieben werden.

\section{Das n"ahere Vorgehen Plancks}
Planck st"utzt sich auf Kirchhoff, der ja einwandfrei argumentiert hatte, 
da"s im thermodynamischen Gleichgewicht die spektrale Energieverteilung 
$\rho(\nu,T)$ eine \emph{universelle} Funktion ist, also von der Form 
des Hohlraums und der Beschaffenheit der W"ande g"anzlich unabh"angig ist.  
Die geniale, aber in den meisten Darstellungen wenig hervorgehobene 
Idee Plancks ist nun folgende (vorgetragen in 
\cite{Planck-GW}, Band\,I, Dokument Nr.\,34, pp.\,592-593): 
wegen der Unabh"angigkeit der spektralen Energieverteilung von der 
physikalischen Beschaffenheit der Wand darf man sich 
\emph{zum Zwecke der theoretischen Bestimmung} der 
Funktion $\rho(\nu,T)$ die Wand auch aus einem hypothetischen, der 
theoretischen Beschreibung leicht zug"anglichen Material ersetzt denken.
Dabei ist es ganz unwesentlich, ob dies hypothetische Medium in der 
realen Welt tats"achlich existiert, sondern wesentlich ist nur, da"s es
den bekannten Gesetzen der Physik gen"ugt, also in diesem Sinne 
existieren \emph{k"onnte}. Die Kirchhoffsche "Uberlegung versichert 
dann, da"s die spektrale Energieverteilung, die sich (theoretisch) 
im Hohlraum des hypothetischen Mediums einstellt, dieselbe ist 
wie die im Hohlraum eines tats"achlich existierenden Materials.
  
Planck w"ahlt als hypothetisches Medium eine Art Gitter von kleinen 
elektrischen Ladungen, die mit einer kleinen Feder elastisch an eine 
Ruhelage befestigt sind. Planck nennt diese Gebilde "`Resonatoren"',
denn sie sollen f"ahig sein, kleine Schwingungen mit einer festen 
Frequenz $\nu$ (der sogenannten "`Eigenfrequenz"') auszuf"uhren, 
wenn sie von einer elektromagnetischen Welle dieser Frequenz getroffen 
werden. Dieses sehr vereinfachte Modell einer "`Wand"' ist nun durch 
die damals bekannten Gesetze der Elektrodynamik und Mechanik 
vollst"andig zu erfassen -- ganz im Gegensatz zu einer realistischen 
Wand, deren mikroskopischer Aufbau und vor allem deren komplizierte 
Wechselwirkung mit auftreffenden Lichtstrahlen zum damaligen Zeitpunkt 
noch ganz unverstanden waren. 

Aus der selbstverst"andlichen Bedingung, da"s im thermodynamischen 
Gleichgewicht jeder dieser elementaren Resonatoren genauso viel 
elektromagnetische Energie emittiert wie absorbiert, leitete Planck 
die folgende Bedingung zwischen spektraler Energiedichte $\rho(\nu,T)$ 
und mittlerer Energie $\E(\nu,T)$ eines einzelnen Resonators 
der Schwingungsfrequenz $\nu$ bei der Temperatur $T$ ab:
\begin{equation}
\label{eq:Planck1}
\rho(\nu,T)=\frac{8\pi\nu^2}{c^3}\E(\nu,T)\,.
\end{equation}
Es mu"s hier nochmals betont werden, da"s diese Gleichung eine 
unzweideutige Folge der Gesetze der klassischen Physik (Mechanik 
und Elektrodynamik) ist. H"atte Planck die damals bereits von seinem 
wissenschaftlichen Widersacher Ludwig Boltzmann (1844-1906) 
ausgearbeitete statistische Mechanik akzeptiert, so h"atte er sofort 
einen Ausdruck f"ur $\E(\nu,T)$ angeben k"onnen. Aus dem 
sogenannten "Aquipartitionsgesetz der statistischen Mechanik folgt 
n"amlich, da"s 
\begin{equation}
\label{eq:Aequipartition}
\E(\nu,T)=\frac{R}{N_A}T\,,
\end{equation}
wobei $R$ die sogenannte universelle Gaskonstante ist (durch Messungen 
gut bekannt) und $N_A$ die Avogadro-Zahl, also die Zahl der in einem Mol 
Gas enthaltenen Molek"ule. Er w"are damit zum sogenannten 
Rayleigh-Jeans-Gesetz gelangt:
\begin{equation}
\label{eq:Rayleigh-Jeans}
\rho(\nu,T)= \frac{8\pi\,\nu^2}{c^3}\frac{R}{N_A}T\,,
\end{equation}
das -- obwohl eine ebenso unzweideutige Folge der klassischen Physik -- 
ganz unsinnige Aussagen macht. Zum Beispiel besagt es, da"s bei 
fester Temperatur $T$ die in elektromagnetischen Wellen der Frequenz 
$\nu$ abgestrahlte Energie quadratisch in $\nu$ w"achst, insgesamt also 
unendlich viel Energie abgestrahlt wird, wenn man "uber alle Frequenzen 
summiert. Auch hinsichtlich der Abh"angigkeit von $T$ geht der Ausdruck 
(\ref{eq:Rayleigh-Jeans}) v"ollig fehl. Das direkt proportionale Ansteigen 
der Strahlungsenergie mit der Temperatur h"atte zum Beispiel zur Folge, 
da"s bei jeder Frequenz die Energieabstrahlung bei Raumtemperatur 
-- $T$ etwa gleich 290\,Grad Kelvin -- immerhin noch ein Sechstel der 
Abstrahlung bei der Temperatur von 1700 Grad Kelvin w"are. Letztere 
entspricht etwa der Temperatur schmelzenden (d.h. wei"sgl"uhenden) 
Stahls. Dies ist offensichtlich eine groteske "Ubersch"atzung der 
Abstrahlung bei Raumtemperatur. Doch Planck erw"ahnt diese 
katastrophale Folge mit keinem Wort. Erst Einstein wird in seiner 
Nobelpreisarbeit von 1905 darauf beharren, da"s die klassische Physik 
notwendig zum inakzeptablen Rayleigh-Jeans-Gesetz f"uhrt und deswegen 
fundamental nicht richtig sein kann. 

Planck geht v"ollig andere, recht seltsame Wege, um die jetzt noch 
fehlende Funktion $\rho(\nu,T)$ zu bestimmen. In der Annahme 
der Richtigkeit des Wienschen Gesetzes kennt er das Ziel und wei"s 
daher, welchen Ausdruck f"ur $\rho(\nu,T)$ er "`herbeiargumentieren"' 
mu"s, um (\ref{eq:Wien2}) aus (\ref{eq:Planck1}) folgen zu lassen. 
An dieser Stelle bringt er nun den 2.\,Hauptsatz der Thermodynamik ins 
Spiel: Statt die Energie $\E(\nu,T)$ des einzelnen Resonators 
zu bestimmen -- wof"ur er keine direkte Methode hat --, geht er den 
Umweg "uber die Entropie $S(\nu,T)$, denn diese sollte sich aus den 
Forderungen des 2.\,Hauptsatzes ergeben. Aus einer allgemein g"ultigen 
thermodynamischen Relation, nach der die Ableitung der Entropie nach 
der Energie die inverse Temperatur ist (siehe Gleichung 
(\ref{eq:Entropie-Energie-Temp}) im Anhang\,\ref{sec:AnhangPlanck}), 
w"urde sich dann auch die Funktion $\E(\nu,T)$ ergeben. Planck gibt 
dann tats"achlich einen Entropieausdruck an, von dem er zeigen kann, 
da"s er allen Anforderungen des 2.\,Hauptsatzes gen"ugt und der 
direkt zum Wienschen Gesetz f"uhrt. Entgegen seiner obigen Aussagen 
zeigt er aber nicht, da"s dieser Ausdruck eindeutig ist. Es k"onnte 
also durchaus andere, ebenfalls mit dem 2.\,Hauptsatz formal 
vertr"agliche Strahlungsgesetze geben (was sich sp"ater auch als 
tats"achlich gegeben herausstellt).

\section{Der Widerspruch}
\begin{figure}[htb]
\noindent
\centering\epsfig{figure=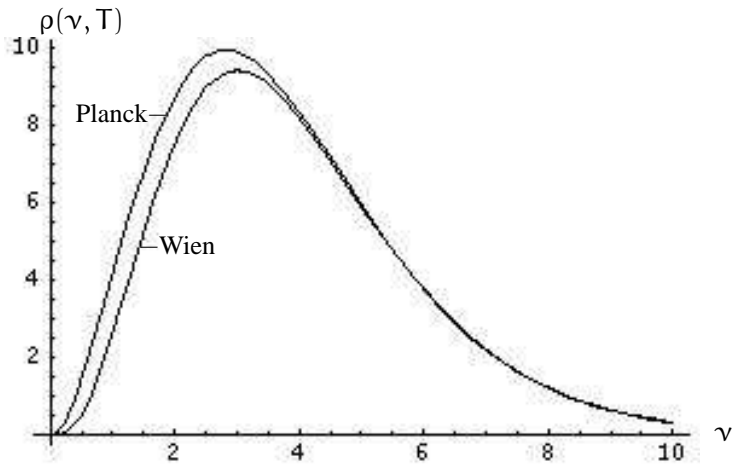, width=0.7\linewidth}
\put(6,9){$\nu$}
\put(-250,165){$\rho(\nu,T)$}
\put(-212,80){\small $-$Wien}
\put(-236,130){\small Planck$-$}
\caption{\small Die spektrale Energieverteilung bei fester 
Temperatur als Funktion der Frequenz nach der Planckschen 
und Wienschen Strahlungsformel. Der bequemeren Darstellbarkeit 
halber sind die Einheiten so gew"ahlt, da"s $a=7$ und 
$b/T=1$.  F"ur kleine Frequenzen (links vom Maximum) 
verl"auft die Plancksche Kurve erkennbar oberhalb der Wienschen, 
w"ahrend sich f"ur gro"se Frequenzen (rechts vom Maximum) 
die Kurven rasch ann"ahern und schlie"slich praktisch zur 
Deckung kommen.}
\label{fig:PlanckWien}
\end{figure}

Experimentelle Messungen an der Physikalisch-Technischen Reichsanstalt in 
Berlin im Jahre 1899 ergaben systematische Abweichungen vom Wienschen 
Strahlungsgesetz im Bereich niederer Frequenzen (d.h. gro"ser Wellenl"angen)
\cite{Lummer-Pringsheim,Rubens-Kurlbaum}.
Die gemessenen Energien lagen bei kleinen Frequenzen systematisch 
oberhalb der Wienschen Kurve. Dazu m"u"sten erst neue Me"smethoden 
entwickelt werden, um den niederfrequenten Anteil des Spektrums m"oglichst 
sauber zu isolieren. Diese "`Divergenzen von erheblicher Natur"' (Planck) 
wurden in der Sitzung der Deutschen Physikalischen Gesellschaft am 
19.\,Oktober mitgeteilt. Es ist bekannt, da"s Planck bereits am 7.\,Oktober 
-- einem Sonntag -- von Heinrich Rubens, einem der Experimentatoren, 
aufgesucht und von den neuen experimentellen Befunden unterrichtet wurde. 
Noch am gleichen Abend fand Planck durch geschicktes Probieren 
(im Anhang\,\ref{sec:AnhangEnergiefluktuationen} erl"autert) eine neue, 
von der Wienschen leicht abweichende Strahlungsformel, die die neuen 
Resultate befriedigend wiederzugeben vermochte. Diese teilte er dann 
ebenfalls am 19.\,Oktober im Anschlu"s an das Referat des 
Experimentalphysikers Kurlbaum der Deutschen Physikalischen Gesellschaft 
mit. Damit war die Plancksche Strahlungsformel geboren:
\begin{equation}
\rho(\nu,T)=\frac{a\nu^3}{\exp\bigl(b\nu/T\bigr)-1}\,.
\label{eq:Planck}
\end{equation}
Sie unterscheidet sich von der Wienschen Formel (\ref{eq:Wien2})
lediglich durch die -1 im Nenner, so da"s f"ur hohe Frequenzen und/oder 
kleine Temperaturen beide Ausdr"ucke approximativ gleich sind.
F"ur kleine Verh"altnisse $\nu/T$ verl"auft die Plancksche Kurve aber 
systematisch \emph{oberhalb} der Wienschen, sagt also bei gegebener 
Temperatur eine merklich h"ohere Energiedichte im Bereich kleiner 
Frequenzen (d.h. gr"o"serer Wellenl"angen) voraus. Dies ist in  
Abbildung\,\ref{fig:PlanckWien} dargestellt. 

Experimentell wurden diese Abweichungen von der Wienschen Formel 
bei gro"sen Wellenl"angen zuerst von Otto Lummer und Ernst 
Pringsheim \cite{Lummer-Pringsheim} gemessen und sofort darauf 
von Heinrich Rubens und Ferdinand Kurlbaum \cite{Rubens-Kurlbaum} 
noch eindr"ucklicher best"atigt. Beide Gruppen arbeiteten zu 
dieser Zeit an der Physikalisch-Technischen Reichsanstalt in 
Berlin-Charlottenburg. Die Me"skurve aus der Originalver"offentlichung 
von Lummer und Pringsheim ist als Abbildung\,\ref{fig:SpekLP} im 
Anhang\,\ref{sec:AnhangLP} wiedergegeben. 

Zum Schlu"s dieses Abschnitts erw"ahnen wir noch, da"s in moderner 
Schreibweise die Konstanten $a$ und $b$ in (\ref{eq:Planck})
durch andere Konstanten ausgedr"uckt werden, n"amlich die 
Lichtgeschwindigkeit $c$, die Boltzmann-Konstante $k=R/N_A$ und das 
Plancksche Wirkungsquantum $h$: 
\begin{equation} 
\label{eq:KonstanteRel}
a=\frac{8\pi h}{c^3}\,,\qquad
b=\frac{h}{k}\,.
\end{equation}
Dieser Zusammenhang wird allerdings erst durch die theoretische 
Begr"undung der Planckschen Strahlungsformel verst"andlich werden.    

\section{Intermezzo: Einsteins Bestimmung der Avogadro-Zahl}
Zu Beginn seiner ber"uhmten Arbeit "uber Lichtquanten aus dem
Jahre 1905 (\cite{Einstein-CW}, Band\,2, Dokument\,14) macht Einstein eine
wichtige Bemerkung, die man etwa so zusammenfassen kann:
Fordert man, da"s das Gesetz (\ref{eq:Rayleigh-Jeans}), was eine 
notwendige Folge der klassischen Physik ist, als Grenzgesetz in der als 
ph"anomenologisch g"ultig angesehenen Planckschen Formel enthalten
ist, so ergibt sich eine von jeder \emph{theoretischen Begr"undung}
der Planckschen Formel \emph{unabh"angige} Methode zur 
Bestimmung der Avogadro-Zahl $N_A$. Entwickelt man die 
Exponentialfunktion im Nenner von (\ref{eq:Planck}) bis zu linearer 
Ordnung in $b\nu/T$, so ergibt sich das Gesetz 
(\ref{eq:Rayleigh-Jeans}) genau dann, wenn die Avogadro-Zahl $N_A$ 
mit den Konstanten $a,b$ des Planckschen Gesetzes in 
folgender Beziehung steht:   
\begin{equation}
N_A=\frac{b}{a}\cdot\frac{8\pi R}{c^3}\,.
\label{Einstein}
\end{equation}
Da $R$ und $c$ gut bekannt sind, liefert jede Bestimmung von $a$ 
und $b$ durch Strahlungsmessungen auch einen Wert f"ur $N_A$. 
Einstein erhielt so den Wert $N_A=6,17\cdot 10^{23}$. Zu dieser Zeit 
war dies der mit Abstand genaueste Wert der Avogadro-Zahl 
(vgl. Kapitel~5 in \cite{Pais}). 

Aber man konnte noch weiter schlie"sen: Aus der Kenntnis der 
Faradaykonstante (elektrische Ladung eines Mols einwertiger Ionen), 
die aus Elektrolysedaten gut bekannt war, erh"alt man nach Division 
durch $N_A$ den Wert der elektrischen Elementarladung $e$. Die 
Elementarladung (Betrag der Ladung eines Elektrons) lie"s sich also 
aus Strahlungsmessungen mit Hilfe der Planckschen Formel gewinnen, 
wobei sich ein weit besserer Wert als jemals zuvor ergab. 
Dies geht eindr"ucklich aus folgendem Vergleich der damals 
diskutierten Werte mit dem heutigen Wert der "`Particle Data Group"' 
(PDG) hervor (in Einheiten von $10^{-10}\,esu$, wo 
$1esu=0{,}1\, Ampere\times meter/c$ die Ladungseinheit 
"`electrostatic units"' bezeichnet):   

\begin{tabular}{rl}
Richarz (1894):         & 1.29 \\
J.J.~Thomson (1898):    & 6.50 \\
Planck/Einstein (1901): & 4.69 \\
PDG (2000):             & 4.803 204 20(19)
\end{tabular}
\medskip

Tats"achlich hatte bereits Planck 1901 die hier angegebenen Werte 
f"ur die Avogadro-Zahl und die Elementarladung mit Hilfe seiner Formel 
und den Ergebnissen von Strahlungsmessungen ausgerechnet
(\cite{Planck-GW}, Band\,I, Dokument Nr.\,44, pp.\,717-727). 
Aber erst Einstein sah, da"s dieses Vorgehen weitgehend unabh"angig 
von Plancks theoretischer Begr"undung seiner Strahlungsformel 
gerechtfertigt werden kann, wenn man nur die Forderung nach dem 
klassischen Limes stellt.\footnote{Dies ist meines Wissens die erste 
Formulierung eines "`Korrespondenzprinzips"', gem"a"s dem die klassische 
Physik in einem geeigneten "`klassischen Limes"' aus der Quantentheorie 
folgen soll. Erst sp"ater hat Niels Bohr diese Forderung zu einem 
allgemeinen Prinzip  erhoben.} Die Ironie dieser Episode ist, da"s 
diese Pr"azisionsbestimmung einer fundamental atomistischen Gr"o"se 
ausgerechnet durch den damaligen Anti-Atomisten Planck erm"oglicht 
wurde.
  
\section{Der "`Akt der Verzweiflung"'}
Wie sollte nun Planck nach all seinen M"uhen, das Wiensche Gesetz 
theoretisch zu begr"unden, eine Ableitung des neuen Gesetzes
(\ref{eq:Planck}) herzaubern? Hatte er nicht noch gerade argumentiert, 
da"s der 2.\,Hauptsatz notwendig zum Wienschen Gesetz f"uhre?
Immerhin blieb er seiner "`klassischen"' Formel (\ref{eq:Planck1}) 
treu und seiner Strategie, die mittlere Resonatorenergie 
$\E(\nu,T)$ aus der Entropie zu bestimmen. Er erkannte jetzt 
endg"ultig, da"s der Ausdruck f"ur letztere, den er vorher nach vielen 
M"uhen erhalten hatte und der ihm scheinbar unausweichlich zum 
Wienschen Gesetz f"uhrte, nicht der formal einzig m"ogliche sein konnte. 
So sehr sich Planck aber auch abm"uhte, eine Begr"undung des 
erforderlichen neuen Ausdrucks zu liefern, es wollte ihm einfach nicht 
gelingen. In seinem Ringen um das Auffinden allgemeiner Methoden, 
die es erlauben w"urden, die Entropie eines Resonators im Strahlungsfeld 
zu berechnen, verfiel er schlie"slich auf den verzweifelten Ausweg, 
ausgerechnet die von ihm bisher vehement bek"ampfte Methode der 
statistischen Interpretation der Entropie seines Widersachers Boltzmann 
zu verwenden. Danach ist die Entropie eine rein kombinatorische Gr"o"se,
die bekannt ist, wenn man die Anzahl der M"oglichkeiten kennt, eine feste 
Energiemenge auf eine feste Anzahl von Resonatoren zu verteilen. 
Diese Anzahl w"are unendlich -- und damit die Entropie unbestimmt --,
wenn jeder Resonator Energie in kontinuierlichen Mengen aufnehmen 
k"onnte. Damit die Entropie endlich herauskommt, mu"s Planck annehmen,
da"s die Gesamtenergie nur in ganzzahligen Vielfachen einer bestimmten 
Grundeinheit "uber die Resonatoren verteilt werden kann. Aus dem 
allgemeinen Gesetz (\ref{eq:Wien1}) ergibt sich, da"s diese 
Grundeinheit proportional zur Eigenfrequenz $\nu$ des Resonators sein 
mu"s. Diese Proportionalit"atskonstante nennt man heute das Plancksche 
Wirkungsquantum $h$. F"ur die Energie-Grundeinheit $\varepsilon$ gilt 
also die Plancksche Formel
\begin{equation}
\label{eq:hnu}
\varepsilon=h\nu\,.  
\end{equation}
F"ur Planck war dies eine rein formale Annahme von h"ochstens 
heuristischer Bedeutung, die er hoffte, sp"ater durch ein physikalisches 
Argument eliminieren zu k"onnen. Immerhin f"uhrte sie ihn zu einer 
Ableitung, die er der Deutschen Physikalischen Gesellschaft in der Sitzung 
am 14.\,Dezember des Jahres 1900 mitteilte. Dieses Datum gilt bis 
heute als die Geburtsstunde der Quantentheorie. "Uber die ihm so 
seltsam aufgezwungene Annahme der Energiequantelung schrieb Planck 
r"uckschauend in einem Brief aus dem Jahre 1931 (\cite{Planck-Brief-1931}):
\begin{quote}
"`Das war eine rein formale Annahme [Energiequantelung], und ich 
dachte mir eigentlich nicht viel dabei, sondern eben nur das, 
da"s ich unter allen Umst"anden, koste es, was es wolle, ein 
positives Resultat herbeif"uhren m"u"ste. [...]
Kurz zusammengefa"st kann ich die ganze Tat als einen Akt 
der Verzweiflung bezeichnen. Denn von Natur bin ich friedlich 
und bedenklichen Abenteuern abgeneigt."'
\end{quote}
Im Anhang\,\ref{sec:AnhangPlanck} ist Plancks "`Akt der Verzweiflung"' 
nochmals etwas genauer beschrieben.

\section{Einsteins Kritik}
Einstein war mit Plancks theoretischer Begr"undung der 
Strahlungsformel (\ref{eq:Planck}) zutiefst unzufrieden. 
Dabei brachte er im wesentlichen zwei Hauptkritikpunkte vor:
\begin{itemize}
\item[1]
Planck benutzt wesentlich Gleichung (\ref{eq:Planck1}), die mit 
Hilfe der Maxwellschen Theorie abgeleitet ist und voraussetzt, da"s 
der Energieaustausch zwischen Planckschen Resonatoren und 
Strahlungsfeld kontinuierlich verl"auft, \emph{im Gegensatz}
zur Quantisierungsannahme (\ref{eq:hnu}). Zwar kann man zun"achst 
argumentieren, da"s (\ref{eq:Planck1}) ja nur f"ur den statistischen 
Mittelwert der Resonatorenergie G"ultigkeit beansprucht und 
somit vielleicht auch unter einem gequantelten Energieaustausch, 
zumindest in guter N"aherung, g"ultig bleibt.\footnote{Dies war 
stets Plancks Haltung, die er noch 1910 "offentlich vertritt; siehe
\cite{Planck-GW}, Band\,2, Dokument\,71.} Doch w"are das nur 
dann zu erwarten, wenn die mittlere Energie des einzelnen 
Resonators $\E(\nu,T)$ sehr gro"s gegen die Energieportionen 
(\ref{eq:hnu}) ist. Aus (\ref{eq:Planck1}) und der Planckschen 
Formel (\ref{eq:Planck}) kann man aber $\E(\nu,T)$ direkt 
ablesen. Mit den Bezeichnungen (\ref{eq:KonstanteRel}) ergibt sich  
\begin{equation}
\label{eq:ResEnergieMittel}
\E(\nu,T)=\frac{h\nu}{\exp(h\nu/kT)-1}\,.
\end{equation}
Demnach ist sogar umgekehrt $\E(\nu,T)$ sehr viel kleiner als 
$h\nu$, falls $h\nu$ viel gr"o"ser als $kT$ ist. Dies ist genau 
im Geltungsbereich des Wienschen Gesetzes der Fall, in dem die 
Annahme von (\ref{eq:Planck1}) mit der Quantisierungsvorschrift 
($h\nu\gg kT$) also unvertr"aglich zu sein scheint.  
\item[2]
Zur Berechnung von $\E(\nu,T)$ "uber die Entropie verwendet Planck die 
Boltzmannsche Entropiedefinition (\ref{eq:Boltzmann-Entropie}). 
Die zun"achst nur durch formales Abz"ahlen bestimmte mikroskopische 
Multiplizit"at eines makroskopischen Zustands ist jedoch nur dann 
proportional seiner physikalischen Wahrscheinlichkeit, wenn die 
Mikrozust"ande im Sinne der tats"achlich gegebenen Dynamik des 
Systems auch \emph{physikalisch gleich wahrscheinlich} sind, 
soll hei"sen: im Laufe einer langen Zeit mit gleichen relativen 
Zeitdauern eingenommen werden. Da Planck f"ur den Resonator die 
klassische Dynamik als richtig annimmt (z.B. in der Ableitung der 
Gleichung (\ref{eq:Planck1})), w"urde die Boltzmannsche 
Entropiedefinition korrekt angewendet notwendig zur 
Rayleigh-Jeans-Formel (\ref{eq:Rayleigh-Jeans}) f"uhren. 
Diesbez"uglich kommentiert Einstein 1909 in der ihm eigenen 
charmant-frechen Weise 
(\cite{Einstein-CW}, Band\,2, Dokument\,56, pp.\,544-545):  
\begin{quote}
"`So sehr sich jeder Physiker dar"uber freuen mu"s, da"s sich Herr 
Planck in so gl"ucklicher Weise "uber diese Forderung hinwegsetzte,
so wenig w"are es angebracht, zu vergessen, da"s die Plancksche 
Strahlungsformel mit der theoretischen Grundlage, von welcher Herr 
Planck ausgegangen ist, unvereinbar ist."'
\end{quote}
\end{itemize}
 
Diese Bedenken tr"agt Einstein u.a. auch in seinem umfassenden 
Bericht "`"Uber die Entwicklung unserer Anschauungen "uber das 
Wesen und die Konstitution der Strahlung"' auf der 81. Versammlung 
Deutscher Naturforscher und "Arzte 1909 in Salzburg vor, wo der 
30-J"ahrige seinen ersten gr"o"seren "offentlichen Auftritt hatte. 
In der sich anschlie"senden Diskussion erl"autert Planck nochmals seine 
Sichtweise der Quantisierungsannahme (\ref{eq:hnu}), die er dezidiert 
aufgefa"st wissen wollte als Ausdruck eines noch unverstandenen 
Mechanismus, der lediglich die \emph{Wechselwirkung} von Strahlung und 
Materie betraf. Materie war eben nur in der Lage, Energie in gewissen 
endlichen Portionen an das Strahlungsfeld anzugeben oder aus dem 
Strahlungsfeld aufzunehmen. Weder hatte Planck im Sinn, damit eine 
grunds"atzliche Modifikation der Dynamik des Resonators 
auszusprechen und schon gar nicht eine Quantisierung des 
Strahlungsfeldes selbst zu postulieren. Er hoffte auf jeden Fall, 
die Maxwellsche Theorie des Elektromagnetismus, die durchweg von 
der Vorstellung kontinuierlicher Prozesse in Raum und Zeit ausgeht, 
zumindest im wechselwirkungsfreien Fall beizubehalten. 
W"ortlich sagte Planck: (\cite{Einstein-CW}, Band\,2, Dokument 
Nr.\,61, pp.\,585-586):
\begin{quote}
"`Jedenfalls meine ich, man m"u"ste zun"achst versuchen,
die ganze Schwierigkeit der Quantentheorie zu verlegen 
in das Gebiet der \emph{Wechselwirkung} zwischen der 
Materie und der strahlenden Energie; die Vorg"ange im 
reinen Vakuum k"onnte man dann vorl"aufig noch mit den 
Maxwellschen Gleichungen erkl"aren."'
\end{quote}
Dem gegen"uber steht Einsteins Resumee seines Vortrages: 
(\cite{Einstein-CW}, Band\,2, Dokument Nr.\,60, pp.\,576-577):
\begin{quote}
"`Die Plancksche Theorie annehmen hei"st nach meiner Meinung 
geradezu die Grundlagen unserer Strahlungstheorie verwerfen."'
\end{quote}

W"ahrend Plancks ablehnende Haltung gegen"uber einer Modifikation 
der Maxwellschen Theorie des freien Strahlungsfeldes aus seinen 
Schriften ganz offenbar wird (was sich auch in seiner Kritik der 
Lichtquantenhypothese "au"sert), ist seine Haltung gegen"uber einer 
Modifikation der mechanischen Gesetze, hier im Zusammenhang mit den 
Resonatoren, etwas umstritten. Diesbez"uglich 
hat sich in j"ungerer Zeit sogar ein sogenannter "`Historikerstreit"' 
entz"undet (vgl. \cite{Darrigol}), der mir aber etwas "ubertrieben 
scheint. In seiner urspr"unglichen Ableitung macht Planck in der Tat 
die formale Annahme (\ref{eq:hnu}), ohne eine Modifikation der 
mechanischen Gesetze zu erw"ahnen. 1906 gibt Einstein eine Ableitung 
des Planckschen Gesetzes mit Hilfe der von ihm selbst entwickelten 
allgemeinen Methoden der statistischen Mechanik (\cite{Einstein-CW}, 
Band\,2, Dokument\,34). Dort zeigt er, da"s man konsistent (d.h. unter 
Vermeidung der oben unter Punkt\,2 ge"au"serten Kritik) zu 
(\ref{eq:ResEnergieMittel}) gelangt, wenn man annimmt, da"s die 
Resonatoren selbst nur ganzzahlige Vielfache der Energie $h\nu$ 
annehmen k"onnen und formuliert dies als eigentliche, der Planckschen 
Ableitung zugrundeliegende Annahme. Auf die Verallgemeinerung dieser 
Annahme auf jedes schwingungsf"ahige Gebilde in einem Festk"orper 
st"utzt Einstein kurz darauf seine Quantentheorie der spezifischen 
W"arme (\cite{Einstein-CW}, Band\,2, Dokument\,38). Planck ist 
damit aber nicht einverstanden und versucht sp"ater (1911-12) sogar, 
eine Ableitung seines Strahlungsgesetzes zu geben, in der nur der 
Proze"s der Emission, nicht jedoch der Proze"s der Absorption 
"`gequantelt"' ist (\cite{Planck-GW}, Band\,2, 
Dokumente\,73,74,75).\footnote{Dadurch erh"alt er eine Modifikation 
seines fr"uheren Ausdrucks (\ref{eq:ResEnergieMittel}) f"ur die 
mittlere Energie eines Resonators um einen additiven Term $h\nu/2$. 
Dies markiert das erste Auftreten der heute aus der Quantenmechanik 
wohlbekannten "`Nullpunktsenergie"'.} Die Unterscheidung dieser 
"`neuen Strahlungshypothese"' Plancks von der urspr"unglichen ist 
aber nur dann sinnvoll, wenn man annimmt, da"s die Resonatorenergien 
grunds"atzlich kontinuierliche Werte annehmen k"onnen. Daraus mu"s 
man m.E. schlie"sen, da"s zwar Einstein, aber nicht Planck die 
Gleichung (\ref{eq:hnu}) im heutigen quantenmechanischen Sinne 
verstanden haben wollte, n"amlich als allgemeine 
Quantisierungsbedingung materieller schwingungsf"ahiger Systeme. 
In seiner "`Lichtquantenhypothese"'  erweiterte Einstein diese 
Quantisierungsbedingung dann auch auf das freie Strahlungsfeld, 
was der heutigen Sichtweise der Quantenelektrodynamik entspricht.       

\section{Einsteins Lichtquantenhypothese}
In den uns vorliegenden schriftlichen Dokumenten Einsteins kennzeichnet 
er nur eine einzige seiner wissenschaftlichen Ideen als "`sehr revolution"ar"'
(\cite{Einstein-CW}, Band\,5, Dokument Nr.\,27, p.\,31)\footnote{Bei 
diesem Dokument handelt es sich um einen Brief Einsteins an seinen 
Freund Conrad Habicht vom Mai 1905, dem Einstein vier wissenschaftliche 
Arbeiten mit folgenden Worten ank"undigt:
"`Ich verspreche Ihnen vier Arbeiten daf"ur, von denen ich die erste 
in B"alde schicken k"onnte, da ich die Freiexemplare baldigst erhalten 
werde. Sie handelt "uber die Strahlung und die energetischen 
Eigenschaften des Lichtes und ist sehr revolution"ar, wie Sie sehen 
werden, wenn Sie mir Ihre Arbeit \emph{vorher} schicken. [...]
Die vierte Arbeit liegt erst im Konzept vor und ist eine Elektrodynamik 
bewegter K"orper unter Ben"utzung einer Modifikation der Lehre von Raum 
und Zeit; der rein kinematische Teil dieser Arbeit wird Sie interessieren"'.
Die zuletzt, eher lapidar angek"undigte Arbeit, ist die spezielle 
Relativit"atstheorie.}, n"amlich die Lichtquantenhypothese. Diese 
ver"offentlicht er im Jahre 1905, 26-j"ahrig, im gleichen 
Zeitschriftenband wie seine spezielle Relativit"atstheorie und die 
Theorie der Brownschen Bewegung. Die Arbeit tr"agt den Titel 
"`"Uber einen die Erzeugung und Verwandlung des Lichtes betreffenden 
heuristischen Gesichtspunkt"'. Dieser "`heuristische Gesichtspunkt"' 
besteht in einer v"ollig anderen Interpretation der Planckschen 
Quantisierungsbedingung (\ref{eq:hnu}), n"amlich als Eigenschaft des 
Strahlungsfeldes selbst. 
Er schreibt (\cite{Einstein-CW}, Band\,2, Dokument Nr.\,14, p.\,151):
\begin{quote}
"`Nach der hier ins Auge zu fassenden Annahme ist bei Ausbreitung
eines von einem Punkte ausgehenden Lichtstrahles die Energie nicht 
kontinuierlich auf gr"o"ser und gr"o"ser werdende verteilt, sondern 
es besteht dieselbe aus einer endlichen Zahl von in Raumpunkten 
lokalisierten Energiequanten, welche sich bewegen, ohne sich zu 
teilen und nur als Ganze absorbiert und erzeugt werden k"onnen."' 
\end{quote}

Diese scheinbare R"uckkehr zur l"angst "uberkommenen Partikelvorstellung 
des Lichts, die zwar noch Newton vertreten hatte, die aber dann im fr"uhen 
19.\,Jahrhundert durch den Siegeszug der Wellentheorie geradezu 
hinweggefegt wurde, mu"ste auf die Zeitgenossen Einsteins als eine 
Mischung aus naiv und provokant gewirkt haben, eben geradezu unverfroren. 
"Au"serungen dazu werden uns weiter unten begegnen. Und doch war 
Einsteins Sichtweise, wie die seiner Gegner, nicht unbegr"undet. 
Durch eine scharfe Analyse des Strahlungsgesetzes, insbesondere des 
ihm immer suspekt erschienenen Wienschen Bereichs, zeigt er, da"s 
(\cite{Einstein-CW}, Band\,2, Dokument Nr.\,14, p.\,161):
\begin{quote} 
"`Monochromatische Strahlung von geringer Dichte (innerhalb des 
G"ultigkeitsbereiches der Wienschen Strahlungsformel) verh"alt sich
in w"armetheoretischer Beziehung so, wie wenn sie aus voneinander 
unabh"angigen Energiequanten von der Gr"o"se $h\nu$
best"unde"'.\label{zit:EinsteinLQ}
\end{quote}
Die genauere Argumentation Einsteins ist in 
Anhang\,\ref{sec:AnhangEinstein} erl"autert. 

Einstein ist klar, da"s sich diese Vorstellung auch an der Erkl"arung 
bekannter Ph"anomene wird behaupten m"ussen, namentlich solcher, die 
die noch unverstandenen Prozesse bei der Wechselwirkung 
von Licht mit Materie betreffen. Einer dieser Prozesse ist der 
sogenannte "`Photoelektrische Effekt"', bei dem durch Bestrahlung 
einer Metallplatte mit Licht Elektronen aus dem Material 
herausgel"ost werden. Die Energie des ankommenden Lichtes wird 
also durch irgendeinen Proze"s dazu verwandt, das Elektron aus dem 
Atomverband herauszul"osen, wozu eine nur vom Material abh"angige 
Energie $P$ aufzuwenden ist. Die "ubersch"ussige Energie des ankommenden 
Lichtes wird dann in die Bewegungsenergie $E_{\text{kin}}$ des 
austretenden Elektrons investiert. Gem"a"s der traditionellen 
Wellentheorie des Lichtes erfolgt dessen Ausbreitung stetig "uber 
alle Raumbereiche. Da die Energie des Lichtes dann proportional zu 
seiner Intensit"at ist, m"u"ste z.B. die Energie der herausgel"osten 
Elektronen mit dem Abstand der Lichtquelle von der Metallplatte 
fallen, da mit dem Abstand auch die Intensit"at abnimmt. Was aber 
durch den Experimentalphysiker Philipp Lenard 
(1862-1947, Nobelpreis 1905) im Jahre 1900 tats"achlich beobachtet 
wurde, ist, da"s zwar die Anzahl der herausgel"osten Elektronen mit 
fallender Intensit"at abnimmt, nicht aber deren individuelle Energien,
die sich als \emph{von der Intensit"at des eingestrahlten Lichtes 
unabh"angig} ergaben. Auf das einzelne Elektron wird also eine 
immer gleiche Energie "ubertragen. Dieser Tatbestand pa"st nun 
"uberhaupt nicht zur Wellentheorie des Lichtes, wird aber sofort 
plausibel bei Zugrundelegung der Lichtquantenhypothese. Nach dieser 
wird jedes der einzelnen Elektronen durch ein ganzes, unteilbares 
Lichtquant der Energie $h\nu$ herausgel"ost und mit einer 
Bewegungsenergie $E_{\text{kin}}$ heraustreten, die der Differenz 
der Energie des Lichtquants zur Abl"osungsenergie $P$ entspricht:
\begin{equation}
\label{eq:Photoeffekt}
E_{\text{kin}}=h\nu-P.
\end{equation}
Diese sogenannte "`Einsteinsche Gleichung"' zum Photoeffekt wurde 
teilweise durch Lenard und sp"ater vor allem durch den amerikanischen 
Experimentalphysiker Robert Millikan (1868-1953, Nobelpreis 1923)
vollauf best"atigt, was sogar mit ein Grund f"ur die Vergabe des 
Nobelpreises war: "`for his work on the elementary charge of 
electricity and on the photoelectric effect"'. Somit schien der 
Photoeffekt mit einem Schlag eine v"ollig nat"urliche Erkl"arung 
zu finden -- vorausgesetzt, man akzeptierte die 
Lichtquantenhypothese!   

\section{Kritik an der Lichtquantenhypothese}
Trotzdem war aber allen Beteiligten klar, da"s die Annahme der 
Einsteinschen Vorstellung der Lichtquanten v"ollig unvereinbar sein 
w"urde mit der g"angigen (Maxwellschen) Theorie des Elektromagnetismus, 
die wenige Jahre zuvor durch die Aufsehen erregenden Versuche 
Heinrich Hertz' scheinbar so gl"anzend best"atigt wurde und auf die 
Planck die Ableitung seiner Ausgangsgleichung (\ref{eq:Planck1}) 
wesentlich gest"utzt hatte. Die aus der Planckschen Strahlungsformel in 
gewisser Weise ableitbare Lichtquantenhypothese anzunehmen, hie"se dann 
also gleichzeitig, der theoretischen Begr"undung dieser Formel den Boden 
zu entziehen. Das genau war die Kritik Einsteins, die er "uber viele Jahre 
hinweg in mannigfacher Variation immer wieder vorbrachte. Wenig 
"uberraschend ist es daher, da"s die Einsteinsche Lichtquantenhypothese 
vor allem bei Planck, aber auch bei anderen Physikern auf starke 
Ablehnung stie"s, darunter auch solche, die Einstein 
wissenschaftlich und pers"onlich sehr nahe standen (worunter man sonst 
auch Planck z"ahlen mu"s, aber eben mit Ausnahme dieses einen Punktes 
betreffend die Lichtquantenhypothese). So beginnt z.B. der Theoretiker 
Arnold Sommerfeld, einer der besten Kenner der Materie und von Einstein 
sehr geachtet, im Jahre 1911 seinen l"angeren Vortrag auf der 83. 
Versammlung der Gesellschaft Deutscher Naturforscher und "Arzte 
so (\cite{Sommerfeld:1911}, p.\,31)
:%
\begin{quote}
"`Als der wissenschaftliche Ausschu"s unserer Gesellschaft an mich 
die Aufforderung richtete, dieser Versammlung einen Bericht "uber 
die Relativit"atstheorie zu erstatten, erlaubte ich mir dagegen 
geltend zu machen, da"s das Relativit"atsprinzip kaum mehr zu den 
eigentlich aktuellen Fragen der Physik geh"ore. Obwohl erst 6 Jahre
alt -- Einsteins Arbeit erschien 1905 -- scheint es schon in den 
gesicherten Besitz der Physik "ubergegangen zu sein. 
Ganz anders aktuell und problematisch ist die Theorie der 
\emph{Energiequanten} [...]. Hier sind die Grundbegriffe noch im Flu"s 
und die Probleme ungez"ahlt."' 
\end{quote}
Und f"ahrt kurz darauf fort:
\begin{quote}
"`Einstein zog aus der Planckschen Entdeckung die weitestgehenden
Folgen [...] und "ubertrug das Quantenhafte von dem Emissions- 
und Absorptionsvorgang auf die Struktur der Lichtenergie im Raume,
ohne, wie ich glaube, seinen damaligen Standpunkt heute noch 
in seiner ganzen K"uhnheit aufrecht zu erhalten."'
\end{quote}
Und selbst der bereits erw"ahnte gro"se amerikanische 
Experimentalphysiker Robert Millikan, der 10 Jahre seines 
Forscherlebens der experimentellen "Uberpr"ufung der Einsteinschen 
Formel (\ref{eq:Photoeffekt}) f"ur den Photoeffekt widmete und 
dadurch auch die ersten Pr"azisionsmessungen des Planckschen 
Wirkungsquantums $h$ realisierte (siehe Anhang\,\ref{sec:AnhangMillikan}), 
schrieb 1916 in einem langen, zusammenfassenden Artikel "uber die 
gerade von ihm so gl"anzend best"atigte Einsteinsche 
Formel (\cite{Millikan:1916}, p.\,384):
\begin{quote}
"`Despite then the apparently complete success of the Einstein 
equation, the physical theory of which it was designed to be the 
symbolic expression is found so untenable that Einstein himself, 
I believe, no longer holds to it."' 
\end{quote}
Die Lichtquantenhypothese im allgemeinen kommentiert Millikan  
bereits auf der ersten Seite seines Artikels wie folgt
(\cite{Millikan:1916}, p.\,355):
\begin{quote}
"`This hypothesis may well be called reckless, first because an 
electromagnetic disturbance which remains localized in space 
seems a violation of the very conception of an electromagnetic 
disturbance, and second because it flies in the face of the 
thoroughly established facts of interference."' 
\end{quote}

Kurz zuvor, im Jahre 1913, als Einstein die Ehre zuteil wird, 
in die Preu"sische Akademie der Wissenschaften aufgenommen zu werden, 
verfassen Planck, Nernst, Rubens und Warburg ein Empfehlungsschreiben, 
das mit folgenden Worten endet (\cite{Einstein-CW}, Band\,5, 
Dokument Nr.\, 445, p.\,527):
\begin{quote}
"`Zusammenfassend kann man sagen, da"s es unter den gro"sen 
Problemen, an denen die moderne Physik so reich ist, kaum
eines gibt, zu dem nicht Einstein in bemerkenswerter Weise 
Stellung genommen h"atte. Da"s er in seinen Spekulationen
gelegentlich auch einmal "uber das Ziel hinausgeschossen 
haben mag, wie z.B. in seiner Hypothese der Lichtquanten,
wird man ihm nicht allzuschwer anrechnen d"urfen; denn ohne
ein Risiko zu wagen, l"a"st sich auch in der exaktesten 
Naturwissenschaft keinerlei wirkliche Neuerung einf"uhren."'
\end{quote} 
Acht Jahre sp"ater, 1921, wird Einstein f"ur die Erkl"arung des 
Photoelektrischen Effektes mit Hilfe der Lichtquantenhypothese 
der Nobelpreis f"ur Physik zuerkannt. Aber auch danach verklingen 
die Zweifel noch nicht. Ein Jahr nach Einstein bekommt Niels 
Bohr den Nobelpreis. In seiner Nobel-Vorlesung mit dem Titel 
"`The Structure of the Atom"' schl"agt Bohr ganz "ahnliche 
T"one an wie sechs Jahre zuvor Millikan. Unter anderem findet 
sich in der Niederschrift von Bohrs Vorlesung folgender 
eindr"ucklicher Passus~\cite{Bohr-Nobel}: 
\begin{quote}
"`This phenomenon [des Photoelektrischen Effekts], which had been 
entirely unexplainable on the classical theory, was thereby placed 
in quite a different light, and the predictions of Einstein's 
theory have received such exact experimental confirmation in recent 
years, that perhaps the most exact determination of Planck's 
constant is afforded by measurements on the photoelectric effect. 
In spite of this heuristic value, however, the hypothesis of 
light-quanta, which is irreconcilable with so-called interference 
phenomena, is not able to throw light on the nature of radiation. 
I need only recall that these interference phenomena constitute 
our only means of investigating the properties of radiation and 
therefore of assigning any closer meaning to the frequency which 
in Einstein's theory fixes the magnitude of the light-quantum."'
\end{quote}

Wie bereits erw"ahnt, folgte auf Bohr Millikan als 
Physik-Nobelpreistr"ager des Jahres 1923. In seiner 
Nobel-Vorlesung mit dem Titel "` The electron and the light-quant 
from the experimental point of view"' "au"sert auch er sich 
nochmals kritisch, wenn auch mittlerweile in etwas abgeschw"achter 
Form (die Hervorhebungen sind seine)~\cite{MillikanNobel}: 
\begin{quote}
"`
In view of this methods and experiments the general validity 
of Einstein's equation [gemeint ist Gleichung (\ref{eq:Photoeffekt})]
is, I think, now universally conceded, and \emph{to that extent
the reality of Einstein's light-quanta may be considered as
experimentally established.} But the conception of \emph{localized}
light-quanta out of which Einstein got his equation must still be 
regarded as far from being established. Whether the mechanism of 
interaction between ether waves and electrons has its seat in the 
unknown conditions and laws existing within the atom, or is to be 
looked for primarily in the essentially corpuscular 
Thomas-Planck-Einstein conception as to the nature of radiant energy 
is the all-absorbing uncertainty upon the frontiers of modern Physics."'
\end{quote}  

Ein letztes dramatisches Aufb"aumen der Kritiker "au"serte sich 
1924 in einer damals sehr viel Aufsehen erregenden Arbeit von Bohr, 
Kramers und Slater \cite{BKS}, in der eine statistische Theorie der 
Wechselwirkung zwischen Strahlung und Materie formuliert 
wird mit dem erkl"arten Ziel, g"anzlich ohne die Lichtquanten 
auszukommen. Als Preis daf"ur sollte man hinnehmen, da"s die 
Erhaltungss"atze von Energie und Impuls zwar im statistischen 
Mittel, nicht jedoch f"ur den individuellen Elementarproze"s 
g"ultig seien. Dabei hatte gerade ein Jahr zuvor Arthur Compton 
(1892-1962, geteilter Nobelpreis 1927) die klassisch 
unverst"andlichen\footnote{Man beobachtet z.B. eine 
Zunahme der Wellenl"ange des gestreuten R"ontgenlichts, ganz im 
Gegensatz zur wellentheoretischen Streutheorie (nach J.J.~Thomson).   
Im Bild der Lichtquanten entspricht diese einfach der Abgabe von 
Energie des Lichtquants an das als ruhend (bzw. hinreichend 
langsam) angenommene Elektron.} Eigenschaften der Streuung von  
R"ontgenstrahlen an Materie mit der Annahme erkl"art, da"s es 
sich dabei um individuelle St"o"se von Lichtquanten mit einzelnen 
Elektronen handle, wobei f"ur jeden Sto"s individuell Energie- 
und Impulserhaltung gelten~\cite{Compton1} (sogenannter 
Comptoneffekt). Dies mag andeuten, wie verzweifelt der Vorschlag 
von Bohr, Kramers und Slater damals war, die nun argumentierten 
mu"sten, da"s die Ph"anomene auch mit einer nur im statistischen Mittel 
g"ultigen Energieerhaltung vertr"aglich w"aren, was aber schon kurz 
darauf durch zahlreichen Experimente widerlegt wurde 
(z.B. auch wieder durch Compton; siehe \cite{Compton2}). 
Erst ab 1925, das auch das Geburtsjahr der Quantenmechanik ist, 
kann man also davon sprechen, da"s sich Einsteins 
Lichtquantenhypothese in den ma"sgebenden Fachkreisen wirklich 
durchgesetzt hatte.

\section{Zusammenfassung und Ausblick}
Plancks gr"o"ste wissenschaftliche Leistung ist auf ironisch und 
fast tragische Weise erkauft mit dem Scheitern seines gro"s angelegten 
Planes, dessen Ziel es war, den 2.\,Hauptsatz als streng kausales 
Gesetz aus den Gesetzen der Elektrodynamik zu begr"unden. Auf seinem 
Weg dorthin findet er stattdessen ein neues, experimentell gl"anzend 
best"atigtes Strahlungsgesetz unter Zugrundelegung der von ihm sonst 
vehement bek"ampften statistischen Entropiedefinition. Die theoretischen 
Implikationen dieses Gesetzes, namentlich die Lichtquantenhypothese 
Einsteins, entziehen Planck geradezu die gesamte Grundlage, von der aus 
er urspr"unglich startete. Wie kein anderer f"orderte Einstein in dieser 
Zeit durch hartn"ackiges Hinterfragen der Grundlagen der Planckschen 
Strahlungstheorie den endg"ultigen Bruch mit der klassischen Physik. 
So wurde Planck \emph{durch Einstein} zum Revolution"ar wider Willen. 

Aber auch Einstein selbst bleibt dieses Schicksal nicht erspart. 
Noch 1916 gibt er eine wunderbar einfache Ableitung der Planckschen 
Strahlungsformel, die nun vollst"andig auf den Gebrauch der Beziehung 
(\ref{eq:Planck1}) verzichtet.\footnote{Diese Beziehung, die von Planck 
auf rein klassischem Wege abgeleitet wurde, kann tats"achlich auch 
durch die Quantenmechanik und Quantenelektrodynamik begr"undet werden; 
siehe z.B. Kap.\,15 in \cite{Grawert} f"ur eine instruktive 
"`halbklassische"' Ableitung. Im  wesentlichen mu"s das Verh"altnis 
der Wahrscheinlichkeiten f"ur die spontane und induzierte Emission 
berechnet werden.}. Dazu betrachtet Einstein die Absorption und 
Emission von Licht als statistische Prozesse, m"oglicherweise in der 
Hoffnung, sie sp"ater doch noch deterministisch verstehen zu k"onnen. 
Interessanterweise mu"s er, um zur Planckschen Formel zu gelangen, 
neben dem Proze"s der \emph{spontanen} Emission auch einen bis dahin 
unbekannten Proze"s der \emph{induzierten} Emission postulieren (ohne 
ihn w"are er formal zum Wienschen Gesetz gelangt), der sp"ater die 
Grundlage des Funktionsprinzips des Lasers werden sollte. Bez"uglich der 
statistischen Natur dieser Prozesse schreibt er am Ende dieser Arbeit 
(\cite{Einstein-CW}, Band\,6, Dokument Nr.\,38, p.\,396): 
\begin{quote}
"`Die Schw"ache der Theorie liegt einerseits darin, da"s sie uns dem 
Anschlu"s an die Undulationstheorie [d.h. Wellentheorie] nicht n"aher 
bringt, andererseits darin, da"s sie Zeit und Richtung der 
Elementarprozesse [der Lichtabsorption und Emission] dem `Zufall' 
"uberl"a"st; trotzdem hege ich das volle Vertrauen in die Zuverl"assigkeit 
des eingeschlagenen Weges"'. 
\end{quote}
Doch f"uhrte eben dieser eingeschlagene Weg nach weiteren 10 Jahren 
geradewegs zur heutigen Quantenmechanik (1925-26) und 
Quantenelektrodynamik (1928), die Einstein mit seinen 
wissenschaftlichen Grund"uberzeugungen nicht in Einklang bringen 
konnte -- insbesondere deshalb, weil in ihnen der Zufall als 
irreduzibler Bestandteil der Naturerkl"arung auftritt. 
Doch das genauer zu erl"autern, bed"urfte eines weiteren Vortrags.

Betrachtet man r"uckblickend die fr"uhe Entstehungsgeschichte der 
Quantentheorie, so h"atte sie beim besten Willen ironischer nicht 
sein k"onnen. Wir erinnern uns, da"s ihr Ausgangspunkt die 
gemessenen Abweichungen vom Wienschen Gesetz waren, welches zu diesem 
Zeitpunkt (f"alschlicherweise) als strenge Konsequenz der klassischen 
Physik angenommen wurde, ma"sgeblich durch die Arbeiten von Planck. 
Wie Einstein in seiner Lichtquantenarbeit aber gezeigt hatte, 
repr"asentiert das Wiensche Gesetz gerade den typisch 
quantentheoretischen Teilchenaspekt der Strahlung. Die von Lummer 
und Pringsheim gemessenen Abweichungen vom Wienschen Gesetz liegen 
im langwelligen Bereich, in dem das Rayleigh-Jeans-Gesetz 
ann"ahernd g"ultig ist, das nun tats"achlich eine unabweisbare 
Konsequenz der klassischen Physik ist, wie Einstein ebenfalls 
zeigte, und dem Wellenbild der Strahlung entspricht. 
\emph{Etwas "uberspitzt kann man im Nachhinein also 
sagen, da"s die Quantentheorie aus Messungen klassischer 
Korrekturen an einem gl"ucklich erratenen Quantengesetz 
entstand, das irrt"umlich f"ur ein Gesetz der klassischen 
Physik gehalten wurde.}

Als Ausblick sei zum Schlu"s noch erw"ahnt, da"s nicht nur in 
der Wissenschaft vom Kleinsten, sondern auch in den gr"o"sten uns 
heute zug"anglichen Dimensionen, in der Kosmologie, die Plancksche 
Strahlungsformel eine zentrale Rolle spielt. So ist ja unser 
gesamtes Universum ein einziger Strahlungshohlraum, der erf"ullt ist 
von einer elektromagnetischen Strahlung der Temperatur von etwas unter 
3 Grad Kelvin (etwa -270 Grad Celsius). Diese Strahlung entstand etwa 
30\,000 Jahre nach dem Urknall, als sich aus zun"achst gegenseitig 
ungebundenen Elektronen und Atomkernen stabile Atome bildeten. 
Zu diesem Zeitpunkt betrug die Temperatur etwa $100\,000$ Grad 
Kelvin. Wegen der best"andigen Ausdehnung des Universums k"uhlt 
sich die Strahlung stetig ab und hat zur gegenw"artigen Epoche den
eben genannten Wert. 
\begin{figure}[htb]
\noindent
\centering\epsfig{figure=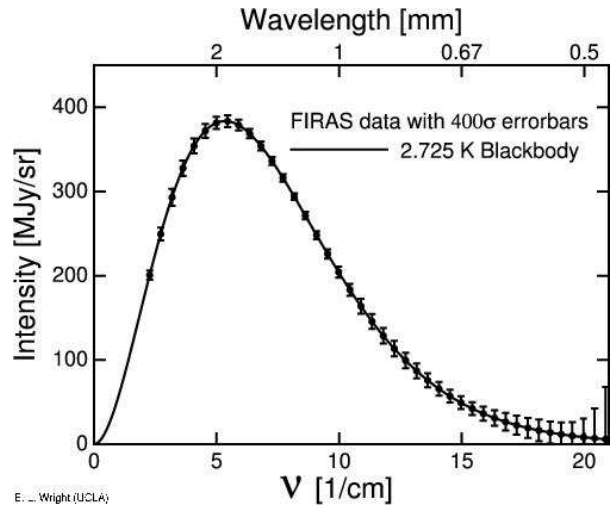, width=0.6\linewidth}
\caption{\small Planck-Spektrum des kosmischen Mikrowellenhintergundes,
aufgenommen durch FIRAS (Far Infrared Absolute Spectrophotometer) des 
Satelliten COBE (Cosmic Background Explorer). Die Fehlerbalken sind 
f"ur 400 Standardabweichungen!}
\label{fig:CMB}
\end{figure}
Seit einigen Jahren werden 
charakteristische Eigenschaften dieser sogenannten "`Kosmischen 
Hintergrundstrahlung"' durch Satelliten vermessen, da diese eine 
reiche F"ulle von Informationen "uber Entwicklung und Zusammensetzung 
unseres physikalischen Universums verraten. Nat"urlich wurde dabei auch 
die spektrale Energieverteilung gemessen und mit der Planckschen 
Formel verglichen. Das Resultat ist in Abbildung\,\ref{fig:CMB} 
dargestellt, in der die Fehlerbalken auf unnat"urliche 400 (!) 
Standardabweichungen vergr"o"sert wurden, damit sie "uberhaupt sichtbar 
sind.  Normale Fehlerbalken von wenigen Standardabweichungen w"aren 
weniger hoch als die Strichdicke der Kurve. Damit ist dies die pr"azisest 
vermessene Planckkurve bis zum heutigen Tag. 
\newpage

\begin{appendix}
\begin{center}
{\Large\textbf{ANH"ANGE}}
\end{center}
\vspace{1.0cm}

\section{N"aheres zu Plancks "`Akt der Verzweiflung"'}
\label{sec:AnhangPlanck}
Es wurde beschrieben, da"s Planck bei der theoretischen Begr"undung 
sowohl des Wienschen als auch seines eigenen Strahlungsgesetzes stets 
von der Beziehung (\ref{eq:Planck1}) ausging und da"s er die darin 
auftretende Funktion $\E(\nu,T)$, die die mittlere Energie eines 
Resonators der Eigenfrequenz $\nu$ im Strahlungsfeld der Temperatur 
$T$ angibt, durch die Entropiefunktion $S(\nu,T)$ dieses Resonators 
zu bestimmen suchte. So ging er auch 
bei der theoretischen Begr"undung seines Gesetzes (\ref{eq:Planck})
am 19.\,Dezember 1900 vor. Dazu wandte er die statistische Definition 
der Entropie von Ludwig Boltzmann an. Diese besagt, da"s die Entropie 
eines Systems proportional zum nat"urlichen Logarithmus des  
statistischen Gewichtes dieses Zustandes ist. Letzteres ist 
definiert als die Anzahl $W$ der M"oglichkeiten, den (makroskopisch 
definierten) Zustand auf verschiedene mikroskopische Arten zu 
realisieren. Dies dr"uckt folgende Formel aus (genannt die 
"`Boltzmannsche"', die aber erst Planck so hinschrieb), die man noch 
heute auf der Grabplatte Boltzmanns auf dem Wiener Zentralfriedhof 
bewundern kann:
\begin{equation}
\label{eq:Boltzmann-Entropie}
S=k\, \ln(W)\,.  
\end{equation}
Dabei ist eben $k$ die Proportionalit"atskonstante zwischen 
Entropie und Logarithmus des statistischen Gewichtes. Man kann 
zeigen, da"s diese Konstante gerade gleich ist dem Quotienten aus 
zwei uns bereits bekannten Gr"o"sen, n"amlich der universellen
Gaskonstante $R$ und der Avogadro-Zahl $N_A$. 

In seiner Verzweiflung, endlich eine theoretische Begr"undung seiner 
bisher nur gl"ucklich erratenen Strahlungsformel (\ref{eq:Planck}) 
liefern zu m"ussen, verfiel Planck auf den Ausweg, die Boltzmannsche 
Gleichung (\ref{eq:Boltzmann-Entropie}) als Definition der Entropie 
zu akzeptieren und sie zur Berechnung der Entropie eines Resonators 
im Strahlungsfeld zu verwenden. Dazu ging Planck so vor:
Angenommen, es gibt $n$ Resonatoren der Eigenfrequenz $\nu$, die 
zusammengenommen in einem Zustand der Energie $E_{\text{total}}$ sind. 
Dann ist das statistische Gewicht $W$ dieses Zustandes definiert 
durch die Anzahl der M"oglichkeiten, die Energie $E_{\text{total}}$ 
auf die $n$ Resonatoren zu verteilen. Physikalisch geht hier die 
oft nicht explizit genannte, aber dennoch sehr wichtige Hypothese 
ein, da"s jede dieser Verteilungen gem"a"s der Dynamik des 
Systems im Laufe der Zeit gleich h"aufig vorkommt.

W"aren die Resonatoren in der Lage, kontinuierliche Mengen von 
Energie aufzunehmen und abzugeben, so w"are das statistische Gewicht 
unendlich und Formel (\ref{eq:Boltzmann-Entropie}) erg"abe ebenfalls 
keinen endlichen Wert. Diesen Schlu"s kann man durch die Annahme umgehen, 
da"s jeder der Resonatoren seine Energie nur portionsweise in 
Einheiten einer festen Grundmenge aufnehmen und abgeben kann. 
Sei diese Grundmenge $\varepsilon$, so gibt es also insgesamt 
$n=E/\varepsilon$ Energieportionen zu verteilen. Es ist nun eine 
elementare kombinatorische Aufgabe, zu berechnen, wie viele 
M"oglichkeiten es gibt, $n$ Portionen Energie auf $N$ Resonatoren
zu verteilen. Die Antwort ist 
\begin{equation}
\label{eq:Kombinatorik}
W=\frac{(n+N-1)!}{n!(N-1)!}\,.  
\end{equation}
Daraus erh"alt man mit (\ref{eq:Boltzmann-Entropie}) die Entropie 
des Zustandes aller Resonatoren der Eigenfrequenz $\nu$ und nach 
weiterer Division durch die Anzahl $N$ dieser Resonatoren die 
gesuchte Entropie eines einzelnen Resonators der Eigenfrequenz 
$\nu$. Das Ergebnis kann man ausdr"ucken\footnote{Man verwendet 
dazu die N"aherungsformel $\ln(N!)\approx N\ln(N)-N$, die f"ur 
gro"se $N$ g"ultig ist. Nach Planck werden an dieser Stelle sowohl 
$N$ als auch $n$ als gro"s angenommen. Letzteres ist tats"achlich 
nicht immer korrekt, was Einstein Planck sp"ater vorwirft.} 
durch die mittlere Energie $E=E_{\text{total}}/N$ eines 
Resonators und des noch unbekannten "`Energiequantums"' 
$\varepsilon$:
\begin{equation}
\label{eq:Entropie-Osz}
S= \bigl(1+E/\varepsilon\bigr)\cdot
\ln\bigl(1+E/\varepsilon\bigr)-
\bigl(E/\varepsilon\bigr)\cdot\ln\bigl(E/\varepsilon\bigr)\,.   
\end{equation}
Damit ist die Aufgabe fast gel"ost. Denn es gilt in der Thermodynamik 
immer (unabh"angig davon, ob man die statistische Interpretation der 
Entropie zugrundelegt), da"s die Ableitung der Entropie nach der 
Energie gleich dem Kehrwert der Temperatur ist:
\begin{equation}
\label{eq:Entropie-Energie-Temp}
\frac{dS}{dE}=\frac{1}{T}.
\end{equation}
Wendet man dies auf (\ref{eq:Entropie-Osz}) an, so kann man sofort 
$E$ als Funktion von $\varepsilon$ und $T$ berechnen, was dann 
eingesetzt in (\ref{eq:Planck1}) f"ur das Strahlungsgesetz liefert:  
\begin{equation}
\label{eq:Osz-Energie}
\rho(\nu, T) =\frac{8\pi\nu^2}{c^3}\frac{\varepsilon}{\exp(\varepsilon/kT)-1}\,.
\end{equation}
Damit dies dann dem Planckschen Strahlungsgesetz (\ref{eq:Planck})
gleicht, mu"s man eine Annahme "uber die tats"achliche Gr"o"se der 
"`Energiequanten"' $\varepsilon$ machen, was ja bisher noch nicht geschehen 
ist. Schon aus einem direkten Vergleich von (\ref{eq:Osz-Energie}) mit 
der allgemein g"ultigen Gleichung (\ref{eq:Wien1}) ergibt sich, da"s 
$\varepsilon$ proportional zu $\nu$ sein mu"s. Nennt man die
Proportionalit"atskonstante $h$, die die Dimension einer Wirkung 
haben mu"s, so hat man gerade (\ref{eq:hnu}), und es ergibt sich 
die Plancksche Formel. 

Denkt man sich die Planckschen Energieportionen als Lichtquanten, 
d.h. im Raum lokalisierte Energiepakete, so entspricht die durch 
(\ref{eq:Kombinatorik}) ausgedr"uckte Abz"ahlung der sogenannten 
\emph{Bose-Einstein-Statistik}. An dieser ist bemerkenswert, 
da"s die Lichtquanten als \emph{ununterscheidbare} Entit"aten 
behandelt werden, d.h. es ist egal, welche der $N$ Lichtquanten 
den individuellen Resonator besetzen, wichtig ist nur ihre Anzahl. 
F"ur Planck war (\ref{eq:Kombinatorik}) jedoch nicht Ausdruck 
einer irgendwie ungew"ohnlichen Statistik, da er nicht im 
Bild der Lichtquanten argumentierte. So bekommt man etwa 
(\ref{eq:Kombinatorik}) auch als Antwort auf die Frage, wieviel 
M"oglichkeiten es gibt, $N$ Kellen Suppe auf $n$ (unterscheidbare) 
Teller zu verteilen. Der Planckschen Quantisierungsannahme 
entspricht hier lediglich die Regel, immer nur ganze Kellen an Suppe 
zu verteilen.

\newpage
\section{N"aheres zu Einsteins Lichtquantenhypothese}
\label{sec:AnhangEinstein}
In diesem Anhang wollen wir etwas n"aher ausf"uhren, durch welche 
mathematische Schlu"skette Einstein zu seiner Lichtquantenhypothese 
gef"uhrt wurde. Grundlage ist wieder das Boltzmannsche Prinzip
(\ref{eq:Boltzmann-Entropie}). In diesem d"urfen wir $W$ auch durch 
die Wahrscheinlichkeit des Makrozustands ersetzen, denn diese 
ist proportional zur Anzahl $W$ seiner mikroskopischen Realisierungen. 
Das Ersetzen von $W$ durch einen dazu proportionalen Ausdruck unter 
dem Logarithmus f"uhrt aber zu einer additiven Konstanten zur Entropie, 
die in den nachfolgenden "Uberlegungen herausf"allt, da stets nur 
Entropie\emph{differenzen} eine Rolle spielen. 

Als Vorbereitung betrachte man ein Gas aus $N$ Atomen in einem 
Volumen $V$ bei fester Temperatur $T$. Hinsichtlich der Dynamik der 
Atome wird nur vorausgesetzt, da"s ihre Aufenthaltswahrscheinlichkeit 
im Volumen konstant ist, da"s also Teilvolumina gleichen Inhalts 
auch mit gleicher Wahrscheinlichkeit von einem Atom besetzt werden. 
Die Wahrscheinlichkeit daf"ur, da"s sich alle Atome in einem Teilvolumen 
$V_0\subset V$ befinden, ist dann gegeben durch $(V_0/V)^N<1$. 
Entsprechend hat dieser Zustand eine um einen Betrag $\Delta S$ 
geringere Entropie als der "uber ganz $V$ gleichverteilte Zustand, wobei 
\begin{equation}
\label{eq:GasEntropie}
\Delta S=S-S_0=k\cdot\ln\left(\frac{V}{V_0}\right)^N\,.   
\end{equation}

Eine analoge "Uberlegung stellt Einstein nun mit W"armestrahlung 
an, ebenfalls im Volumen $V$ bei der Temperatur $T$. Dazu mu"s er 
aber den Ausdruck f"ur die Strahlungsentropie berechnen. 
Diesen erh"alt er so: Sei $\rho(\nu,T)$ die spektrale Dichte der 
Energie (hier als bekannt vorausgesetzt) und $\varphi(\nu,T)$ der 
(zu bestimmende) Ausdruck f"ur die spektrale Dichte der Entropie. 
Das hei"st, da"s der auf das Volumen $V$ und das Frequenzintervall 
$[\nu\,,\,\nu+d\nu]$ entfallende Anteil der Strahlungsenergie 
durch $\rho(\nu,T)V\,d\nu$ und der Anteil der Strahlungsentropie 
durch $\varphi(\nu,T)V\,d\nu$ gegeben ist. Ganz allgemein gilt in 
der Thermodynamik, da"s die Ableitung der Entropie nach der Energie 
das Inverse der absoluten Temperatur $T$ ist. Das gilt auch f"ur die 
spektralen Verteilungen. Also hat man 
\begin{equation}
\label{eq:EinLicht1}
\frac{\partial \rho}{\partial\varphi}=\frac{1}{T}\,.
\end{equation}
Kennt man das Strahlungsgesetz, d.h. die Funktion $\rho(\nu,T)$, so 
kann man damit auf der rechten Seite $1/T$ als Funktion von $\nu$ und 
$\rho$ ausdr"ucken und die Gleichung integrieren, wodurch man 
$\varphi$ als Funktion von $\nu$ und $\rho$ erh"alt. 

Einstein benutzt nun nicht das Plancksche, sondern das Wiensche 
Gesetz (vgl. (\ref{eq:Wien2}), das sich im Grenzfall 
hoher Frequenzen und/oder kleiner Temperaturen aus ersterem ergibt. 
L"ost man dieses nach $1/T$ auf, setzt es auf der rechten Seite von 
(\ref{eq:EinLicht1}) ein und integriert einmal nach $\rho$, so erh"alt 
man 
\begin{equation}
\label{eq:EinLicht2}
\varphi(\nu,T)=-\,\frac{\rho}{b\nu}\cdot
\left[\ln\left(\frac{\rho}{a\nu^3}\right)-1\right]+konst.
\end{equation}
Setzt man f"ur die im Volumen $V$ und Frequenzintervall 
$[\nu\,,\,\nu+d\nu]$ enthaltene Energie $E=\rho V\,d\nu$ und 
Entropie $S=\varphi V\,d\nu$, so kann man dies auch so schreiben: 
\begin{equation}
\label{eq:EinLicht3}
S(E,\nu)=-\,\frac{E}{b\nu}\cdot
\left[\ln\left(\frac{E}{Va\nu^3\,d\nu}\right)-1\right]+konst.
\end{equation}
Betrachtet man bei konstantem $E$ und $\nu$ die Differenz der Entropien 
der Strahlung, einmal im Volumen $V$ und einmal im Volumen $V_0$, so erh"alt 
man
\begin{equation}
\label{eq:EinLicht4}
\Delta S=S-S_0=\frac{E}{b\nu}\cdot\ln\left(\frac{V}{V_0}\right)
=k\cdot\ln\left(\frac{V}{V_0}\right)^{E/b k\nu}\,.
\end{equation}
Dies vergleicht Einstein mit (\ref{eq:GasEntropie}) und kommt zu dem 
Schlu"s, da"s sich W"armestrahlung im G"ultigkeitsbereich des
Wienschen Strahlungsgesetzes entropisch gesehen so verh"alt, wie ein 
Gas aus $N=E/b k\nu$ Atomen (siehe das Zitat Einsteins auf 
Seite\,\pageref{zit:EinsteinLQ}). Die "`Atome"' des Lichts hei"sen 
\emph{Lichtquanten}. Sie sind (im G"ultigkeitsbereich des Wienschen 
Gesetzes!) als r"aumlich lokalisiert zu denken und haben die Energie 

\begin{equation}
\label{eq:EinLicht5}
\varepsilon=E/N=b k\nu=h\nu\,,
\end{equation}
wobei wir noch ausgenutzt haben, da"s die Konstante $b$ des 
Wienschen Gesetzes mit der Planckschen Konstante $h$ "uber die 
Boltzmann-Konstante $k$ gem"a"s $h=bk$ verbunden ist.

\newpage
\section{Erster experimenteller Hinweis auf die Quantentheorie}
\label{sec:AnhangLP}
\begin{figure}[!h]
\centering\epsfig{figure=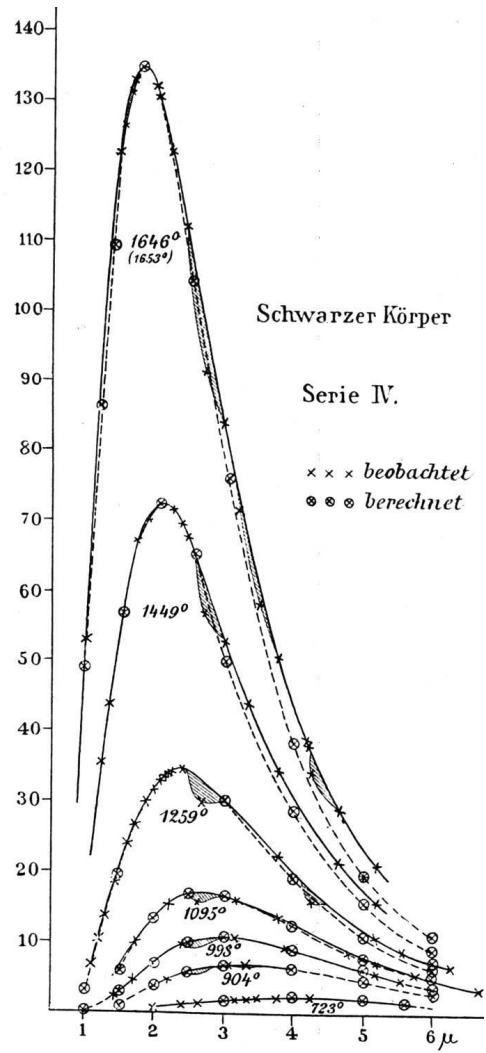, width=0.5\linewidth}
\caption{\label{fig:SpekLP}
Wiedergabe von Fig.\,1 aus der 
Originalver"offentlichung~\cite{Lummer-Pringsheim}.
Aufgetragen nach oben ist die Energiedichte der Strahlung, 
nach rechts ihre Wellenl"ange in Einheiten von $\mu=10^{-6}\,m$.
Die durch die Symbole $\times$ markierte, von Lummer und Pringsheim 
gemessene Kurve, verl"auft im langwelligen Bereich (rechts, jenseits
des Maximums) systematisch oberhalb der durch die Symbole $\otimes$ 
markierten Kurve, die den theoretisch bestimmten Werten gem"a"s 
der Wienschen Strahlungsformel (\ref{eq:Wien2}) entspricht. 
Die schraffierten Einbuchtungen der gemessenen Kurve bei etwa 
$2{,}7\,\mu$ und $4{,}5\,\mu$ werden durch bekannte 
Absorptionsbanden des Wasserdampfes bzw. der Kohlens"aure 
verursacht.}
\end{figure}

\newpage
\section{Millikans Messungen zum Photoeffekt und seine 
Pr"azisionsbestimmung des Planckschen Wirkungsquantums}
\label{sec:AnhangMillikan}
\begin{figure}[!h]
\centering\epsfig{figure=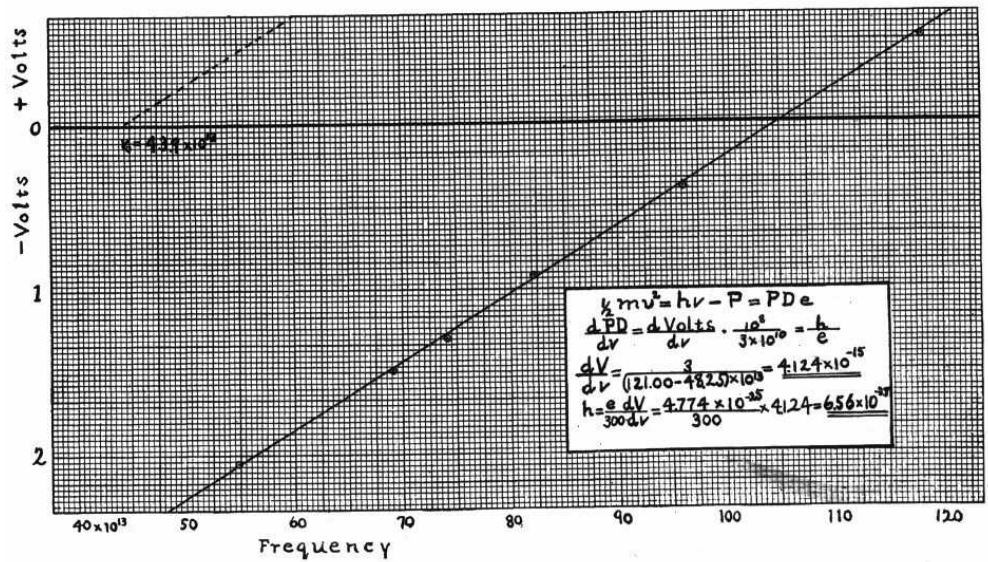, width=1.0\linewidth}
\caption{\label{fig:Millikan}
Me"skurve aus Millikans Ver"offentlichung \cite{Millikan:1916},
in der die lineare Beziehung (\ref{eq:Photoeffekt}) zwischen 
der kinetischen Energie des aus dem Metall gel"osten Elektrons 
und der Frequenz des eingestrahlten Lichts deutlich sichtbar 
ist. Auf der Ordinate ist statt $E_{\text{kin}}$
die Spannung $V=E_{\text{kin}}/e$ ($e$=Elementarladung) in $Volt$ 
aufgetragen, auf der Abszisse die Frequenz des Lichts in Einheiten 
von $10^{13}\,Hz$. Aus der Steigung (und dem Wert f"ur die 
Elementarladung) erh"alt Millikan einen Wert f"ur das Plancksche 
Wirkungsquantum  von $h=6{,}56\cdot 10^{-27}\,erg$ 
($erg=g\,cm^2/s^2=10^{-7}Joule$). Dieser liegt 1\% unterhalb des 
heute genauesten Wertes von $6,629\,0693\cdot 10^{-27}erg$ (relativer 
Fehler $1,7\cdot 10^{-7}$). Gemittelt "uber verschiedene Messungen 
kommt Millikan tats"achlich noch etwas n"aher an diesen Wert. 
Da Millikan die kinetische Energie der Austrittselektronen "uber 
die von ihnen durchlaufene Spannung $V$ mi"st, kommt der Wert der 
Elementarladung $e$ ins Spiel und somit auch dessen experimentelle 
Unsicherheit.  
}
\end{figure}  
\newpage

\section{Energiefluktuationen}
\label{sec:AnhangEnergiefluktuationen}
Als Planck seine Formel zum ersten Mal niederschrieb, geschah dies 
ohne Kenntnis der erst sp"ater von ihm erdachten Ableitung. 
Vielmehr erhielt er sie, indem er eine gewisse thermodynamische 
Gr"o"se f"ur die Wiensche und die Rayleigh-Jeanssche Formel einfach 
addierte. Eine physikalische Interpretation dieses formalen Vorgehens 
hatte Planck nicht -- wie er sp"ater selber zugab. Erst Einstein hat 
diese Interpretation sp"ater geliefert, die einen interessanten 
Aspekt des "`Welle-Teichen-Dualismus"' darstellt.   

Setzen wir zur Abk"urzung $\beta:=1/(kT)$ und sei $\E$ wieder die 
mittlere Energie eines Resonators, jetzt aufgefa"st als Funktion von 
$\beta$ (statt $T$) und $\nu$, so ist die von Planck 
betrachtete Gr"o"se gegeben durch die Ableitung $-\,d\E/d\beta$. 
Was bedeutet Sie? Bevor wir dies kl"aren, wollen wir sie f"ur die 
aus den drei Strahlungsgesetzen (Rayleigh-Jeans, Wien, Planck)  
folgenden Ausdr"ucke f"ur $\E$ berechnen. 
Es ist 
\begin{equation}
\label{eq:Flukt1}
\E=
\begin{cases}
1/\beta                         &\quad\text{Rayleigh-Jeans}\\
h\nu\,\exp(-\beta h\nu)         &\quad\text{Wien} \\
\frac{h\nu}{\exp(\beta h\nu)-1} &\quad\text{Planck}\,,
\end{cases}
\end{equation}
also gilt
\begin{equation}
\label{eq:Flukt2}
-\,\frac{d\E}{d\beta}=
\begin{cases}
\E^2          &\quad\text{Rayleigh-Jeans}\\
h\nu\,\E      &\quad\text{Wien}\\
\E^2+h\nu\,\E &\quad\text{Planck}\,.
\end{cases}
\end{equation}
Somit ist in der Tat f"ur das Plancksche Gesetz diese Gr"o"se 
(als Funktion von $\E$) additiv aus den entsprechenden Ausdr"ucken 
des Rayleigh-Jeansschen und Wienschen Gesetzes zusammengesetzt. 

Einsteins "Uberlegungen sind nun statistischer Natur -- genauer 
gesagt betrachtet er statistische Fluktuationen der Energie des 
Strahlungsfeldes, was man wegen (\ref{eq:Planck1}) auch auf die 
Resonatoren "ubertragen kann. Dabei ist seine zentrale Idee, 
die Boltzmannsche Gleichung (\ref{eq:Boltzmann-Entropie}) umgekehrt 
zu lesen, d.h. das statistische Gewicht als Funktion der Entropie 
auszudr"ucken. Dr"uckt man die Entropie (bei fester Temperatur) 
als Funktion der Energie aus und entwickelt um das dem 
Gleichgewicht entsprechende lokale Maximum bei $\E= \E_0$, 
so kann man daraus in quadratischer Ordnung die 
normierte Wahrscheinlichkeitsverteilung f"ur eine 
Energiefluktuation $\epsilon=\E-\E_0$ ableiten: 
\begin{equation}
\label{eq:Flukt3}     
P(\epsilon)=\sqrt{\frac{\gamma}{2\pi}}\,
\exp\bigl(-\tfrac{1}{2}\gamma\epsilon^2\bigr)\,.
\end{equation}
Hier ist 
\begin{equation}
\label{eq:Flukt4}
\gamma:=-k^{-1}\cdot\frac{d^2}{d\E^2}\Big\vert_{\E=\E_0}
=-\frac{d\beta}{d\E}\Big\vert_{\E=\E_0}\,, 
\end{equation}
wobei die zweite Gleichheit aus der allgemein g"ultigen 
thermodynamischen Relation (\ref{eq:Entropie-Energie-Temp})
folgt. Also ist das mittlere Schwankungsquadrat der Energie 
gegeben durch:
\begin{equation}
\label{eq:Flukt5}
\langle\epsilon^2\rangle:=\int_{-\infty}^\infty 
P(\epsilon)\epsilon^2\,d\epsilon=\gamma^{-1}=-\,\frac{d\E}{d\beta}\,.
\end{equation}

Damit ist die Gr"o"se, die Planck seiner formalen Interpolation
zugrundelegte, als das mittlere Schwankungsquadrat der Energie 
erkannt. Dieses verh"alt sich bei der Planckschen Formel so, 
als ob es zwei statistisch unabh"angige Ursachen h"atte: die 
Energieschwankungen der Rayleigh-Jeans-Formel, die man mit dem 
klassischen Wellenbild erkl"aren kann, und die der Wienschen 
Formel, die dem Teilchenbild der Lichtquanten entspricht. 
In dieser Hinsicht vereinigt die Plancksche Formel beide Aspekte 
in gleichberechtigter Weise. 

Stellt man das Gesagte konsequent im Teilchenbild (Lichtquanten) 
dar, so  kann man statt von Resonatorenergien von Besetzungszahlen
$n$ sprechen, indem man jede Energie durch $h\nu$ dividiert. Aus der 
letzten Zeile in (\ref{eq:Flukt2}) erh"alt man dann einen einfachen 
Ausdruck f"ur das Schwankungsquadrat der Besetzungszahl: 
\begin{equation}
\label{eq:Flukt6}
\langle (n-\bar n)^2\rangle=\bar n +\bar n^2\,.
\end{equation}       
Der erste Term w"are alleine vorhanden, wenn es sich um klassisch 
unabh"angige Teilchen handelte, wie man leicht nachpr"uft.\footnote{Die
Wahrscheinlichkeit, von $N$ unterscheidbaren Teilchen irgendwelche 
$n$ in einem Zustand der Wahrscheinlichkeit $p$ anzutreffen (z.B. 
im Teilvolumen $V_0\subset V$ zu sein, wobei $V_0/V=p$), ist 
$W(n)=\binom{N}{n}p^n(1-p)^{N-n}$. Man berechnet nun leicht 
$\bar{n}=\langle n\rangle:=\bar n=\sum_{n=0}^NnW(n)=Np$ und 
$\langle n(n-1)\rangle:=\sum_{n=0}^Nn(n-1)W(n)=N(N-1)p^2$. 
Aus beiden zusammen ergibt sich $\langle(n-\bar{n})^2\rangle=\bar{n}(1-p)$, 
was f"ur kleine $p$ in $\bar{n}$ "ubergeht.} 

\end{appendix}


\newpage
\begin{thebibliography}{99}
%
\bibitem{Bohr-Nobel}
Niels Bohr:
"`The Structure of the Atom"'.
Nobel-Vorlesung vom 11. Dezember 1922 (Nobelpreis 1922). \\
Online unter
$\langle$nobelprize.org/physics/laureates/1922/bohr-lecture.html$\rangle$.
%
\bibitem{BKS}
Niels Bohr, Hendrik Kramers und J. Slater:
"`"Uber die Quantentheorie der Strahlung"'.
Zeitschrift f"ur Physik, 24 (1924) pp.\,69-87.
%
\bibitem{Compton1}
Arthur H. Compton:
"`A quantum theory of the scattering of X-rays by light elements."'
Physical Review 21 (1923) pp.\,483-502.
%
\bibitem{Compton2}
Arthur H. Compton:
"`Directed quanta of scattered X-rays"'.
Physical Review 26 (1925) pp.\,289-299.
%
\bibitem{Darrigol}
Oliver Darrigol:
"`The historians disagreement over the meaning of Planck's quantum"'. 
Centaurus, 43 (2001) 219-239.\\ 
Online unter 
$\langle$www.mpiwg-berlin.mpg.de/de/forschung/preprints.html$\rangle$
als Preprint Nr.\,150 verf"ugbar. 
%
\bibitem{Einstein-CW}
Albert~Einstein:
Collected Works (Princeton University Press).
Siehe auch die Internetseite des "`Einstein Papers Project"': 
$\langle$www.einstein.caltech.edu$\rangle$.
%
\bibitem{Grawert}
Gerald Grawert:
"`Quantenmechanik"' (Akademische Verlagsgesellschaft Wiesbaden 1977). 
%
\bibitem{Lummer-Pringsheim}
Otto Lummer und Ernst Pringsheim:
"`Die Vertheilung der Energie im Spectrum des schwarzen 
K"orpers und des blanken Platins"'.
Verhandlungen der Deutschen Physikalischen Gesellschaft im 
Jahre 1899, erster Jahrgang, pp.\,23-41. Herausgegeben von 
Arthur K"onig,  Verlag von Johann Ambrosius Barth, Leipzig 1899.
%
\bibitem{Millikan:1916}
Robert Millikan:
"`A Direct Photoelectric Determination of Planck's `$h$'."'
Physical Review, Vol.\,7 (1916), pp.\,355-388.
%
\bibitem{MillikanNobel}
Robert Millikan:
"`The electron and the light-quant from the 
experimental point of view"'. 
Nobel-Vorlesung vom 23. Mai 1924 (Nobelpreis 1923). \\
Online unter
$\langle$nobelprize.org/physics/laureates/1923/millikan-lecture.pdf$\rangle$. 
%
\bibitem{Pais}
Abraham Pais:
"`Raffiniert ist der Herrgott..."'
Albert Einstein. Eine wissenschaftliche Biographie
(Vieweg \& Sohn, Braunschweig, 1986).
%
\bibitem{Planck-GW}
Max~Planck:
Physikalische Abhandlungen und Vortr"age, Bd.~I-III 
(Vieweg \& Sohn, Braunschweig, 1958).
%
\bibitem{Planck-Brief-1931}
Max~Planck:
Brief an Robert Williams Wood von 1931. 
Wiedergegeben in "`Fr"uhgeschichte der Quantentheorie"', p.~31,
von A.~Hermann (Physik Verlag, Mosbach 1969).
%
\bibitem{Pringsheim}
Ernst Pringsheim:
"`Einfache Herleitung des Kirchhoff'schen Gesetzes"'.
Verhandlungen der Deutschen Physikalischen Gesellschaft in 
Jahre 1901, dritter Jahrgang, pp.\,81-84. Herausgegeben von 
Arthur K"onig, Verlag von Johann Ambrosius Barth, Leipzig 1901.
%
\bibitem{Rubens-Kurlbaum}
Heinrich Rubens und Ferdinand Kurlbaum:
"`"Uber die Emission langwelliger W"armestrahlen durch 
den schwarzen K"orper bei verschiedenen Temperaturen"'.
Sitzungsberichte der Preu"sischen Akademie der 
Wissenschaften 1900, Gesamtsitzung vom 25.\,Oktober, 
pp.\,929-941. 
%
\bibitem{Sommerfeld:1911}
Arnold Sommerfeld:
"`Das Plancksche Wirkungsquantum und seine allgemeine 
Bedeutung f"ur die Molek"ulphysik"'.
Verhandlungen der Gesellschaft Deutscher Naturforscher 
und "Arzte, 83. Versammlung zu Karlsruhe 1911, zweiter Teil,
pp.\,31-50.
%
\end{thebibliography}
\end{document}